\pdfoutput=1
\documentclass[fleqn,usenatbib]{mnras}

\usepackage{newtxtext,newtxmath}
\usepackage[T1]{fontenc}
\usepackage{ae,aecompl}

\usepackage{times} 
\usepackage{graphicx}	% Including figure files
\usepackage{amsmath}	% Advanced maths commands
\usepackage{amssymb}	% Extra maths symbols
\usepackage{multicol}   % Multi-column entries in tables
\usepackage{pdfpages}   % Add PDF files
\usepackage[export]{adjustbox} %Fix the position of an image
\usepackage{float}
\usepackage{subcaption}
\newcommand{\kms}{km\,s$^{-1}$} 			    % per km/s
\newcommand{\mum}{$\mu$m}					    % \micron
\newcommand{\msun}{M$_{\sun}$}                  % per sun masses
\newcommand{\Spitzer}{{\slshape{Spitzer}}} 	    % per Space Telescopes
\newcommand{\Herschel}{{\slshape{Herschel}}}    % per Herschel
\newcommand{\WISE}{{\slshape{WISE}}}
\newcommand{\radec}{RA.,Dec.\,(J2000)}
\title[Multifrequency study of HH~137 and HH~138]{Multifrequency study of HH~137 and HH~138: Discovering new knots and molecular outflows with Gemini\thanks{This work is based on observations obtained at the Gemini Observatory, which is operated by the Association of Universities for Research in Astronomy, Inc., under a cooperative agreement with the NSF on behalf of the Gemini partnership: the National Science Foundation (United States), National Research Council (Canada), CONICYT (Chile), Ministerio de Ciencia, Tecnolog\'{i}a e Innovaci\'{o}n Productiva (Argentina), Minist\'{e}rio da Ci\^{e}ncia, Tecnologia e Inova\c{c}\~{a}o (Brazil), and Korea Astronomy and Space Science Institute (Republic of Korea).}\ and APEX\thanks{APEX is a collaboration between the Max-Planck-Institute for Radio Astronomie, the European Southern Observatory, and the Onsala Space Observatory.}}

\author[L.V. Ferrero et al.]{Leticia V. Ferrero$^{1,2}$\thanks{Contact e-mail: lvferrero@unc.edu.ar},
Cristina E. Cappa$^{2,3}$,
Hugo P. Salda\~no$^{1}$
Mercedes G\'omez$^{1,2}$,
\newauthor
M\'onica Rubio$^{4}$,
and Guillermo G\"unthardt$^{1}$
\\
$^{1}$Universidad Nacional de C\'ordoba, Observatorio Astron\'omico de C\'ordoba,  Laprida 854, C\'ordoba X5000BGR, Argentina\\
$^{2}$Consejo Nacional de Investigaciones Cient\'ificas y T\'ecnicas (CONICET), C1033AAJ, Argentina\\
$^{3}$Facultad de Ciencias Astron\'omicas y Geof\'isicas, Universidad Nacional de la Plata, Paseo del Bosque s/n, 1900 La Plata, Argentina\\
$^{4}$Departamento de Astronom\'ia, Universidad de Chile, Casilla 36-D, Santiago, Chile
}

\date{Accepted XXX. Received YYY; in original form ZZZ}

\pubyear{2019}

\begin{document}
\label{firstpage}
\pagerange{\pageref{firstpage}--\pageref{lastpage}}
\maketitle
\begin{abstract} 
We present a multi-wavelength study of two HH objects (137 and 138) that may be associated. We use Gemini H$_2$ (2.12~\mum) and K (2.2~\mum) images, as well as APEX molecular line observations and \Spitzer\ image archives. Several H$_2$ knots, linked to the optical chain of knots of HH~137, are identified in the Gemini and \Spitzer\ 4.5~\mum\ images. New shock excited regions related to the optical knots delineating HH~138 are also reported. In addition, a bright 4.5~\mum\ 0.09~pc-long arc-shaped structure, roughly located mid-way between HH~137 and HH~138, is found to be associated with two \Spitzer\ Class~I/II objects, which are likely to be the exciting stars. These sources are almost coincident with a high-density molecular clump detected in $^{12}$CO(3-2), $^{13}$CO(3-2), C$^{18}$O(3-2), HCO$^+$(3-2) and HCN(3-2) molecular lines with an LTE mass of 36~\msun. The $^{12}$CO(3-2) emission distribution over the observed region reveals molecular material underlying three molecular outflows. Two of them (outflows 1 and 2) are linked to all optical knots of HH~137 and HH~138 and to the H$_2$ and 4.5~\mum\ shock emission knots. In fact, the outflow 2 shows an elongated $^{12}$CO blue lobe that coincides with all the H$_2$ knots of HH 137 which end at a terminal H$_2$ bow shock. We propose a simple scenario that connects the outflows to the dust clumps detected in the region. A third possible outflow is located to the north-east projected towards a secondary weak and cold dust clump.
\end{abstract}


%%%%%%%%%%%%%%%%%%%%%%%%%%%%%%%%%%%%%%%%%%%%%%%%%%

%%%%%%%%%%%%%%%%% BODY OF PAPER %%%%%%%%%%%%%%%%%%

\section{Introduction}

Stellar jets and molecular outflows appear during the initial stages of star-formation when the incipient star$+$disk system begins to eject winds along the rotation axis that interact with the surrounding cloud. The high kinetic energy of jets (comparable to the gravitational binding energy of the core where the stars are forming) has the potential to disperse the entire envelope of the new stars \citep{Tsinganos2009}. Such an interaction can extend to a few parsecs away from the source. The jets can be observed over a wide range of wavelengths from the ultraviolet to the radio. In the optical they are evident as Herbig-Haro (HH) objects, while in the near infrared (NIR) they are observed as H$_2$ (2.12\,\mum) knots \citep{Reipurth-Bally2001,Bally2016}. In particular, the H$_2$ line at 2.12\,\mum\ is a well-known shock tracer, where the relatively fast jet hits and sweeps up the cold molecular cloud environment. This produces the molecular outflows which are detected at radio wavelengths, mainly in the $^{12}$CO and SiO lines \citep{Davis2010,Beuther2002,Arce2007,Maud2015}.

The outflows are a ubiquitous phenomenon. They are present in objects within a wide range of masses from brown dwarfs \citep[e.g.,][]{Riaz2017,Whelan2018} to massive protostars \citep[e.g.,][]{Fedriani2018,McLeod2018}. Such outflows broaden the molecular lines since they increase the motions within the cloud introducing non-Guassian wings into the $^{12}$CO spectra. The outflows occur simultaneously with the infall motion, where the protostars are
forming. The infall motion is revealed by means of double-peaked $^{12}$CO (or HCO$^+$) line profiles with the blue-shifted peak brighter than the red one \citep{Evans1999}. The depression between both sides almost coincides with an optically thin line (such as C$^{18}$O), which frequently shows a single component peaking at the systemic velocity (e.g. \citealt{Zhou_Evans1994,Cunningham2018}).

Jets not only affect the nearby molecular cloud morphologically, but they also heat and compress the gas rapidly, triggering different molecular processes (such as molecular dissociation, sublimation of grain mantle ices, etc.) contributing to the chemical enrichment of the vicinity of young stars \citep{Tsinganos2009}. Thus, jets and outflows have significant consequences in the evolution of the environments of new stars. Although they have been widely studied in the literature, few of them have been observed at high-resolution in the infrared wavelengths and studied in a wide spectral range. Here, we present a new study of the HH objects named as HH\,137 and HH\,138 using infrared high-resolution observations obtained with the Gemini South telescope and radio observations of molecular emissions (CO isotopologues, HCO$^+$, HCN) taken with the APEX telescope.

\begin{figure}
    \centering
    \includegraphics[width=\columnwidth]{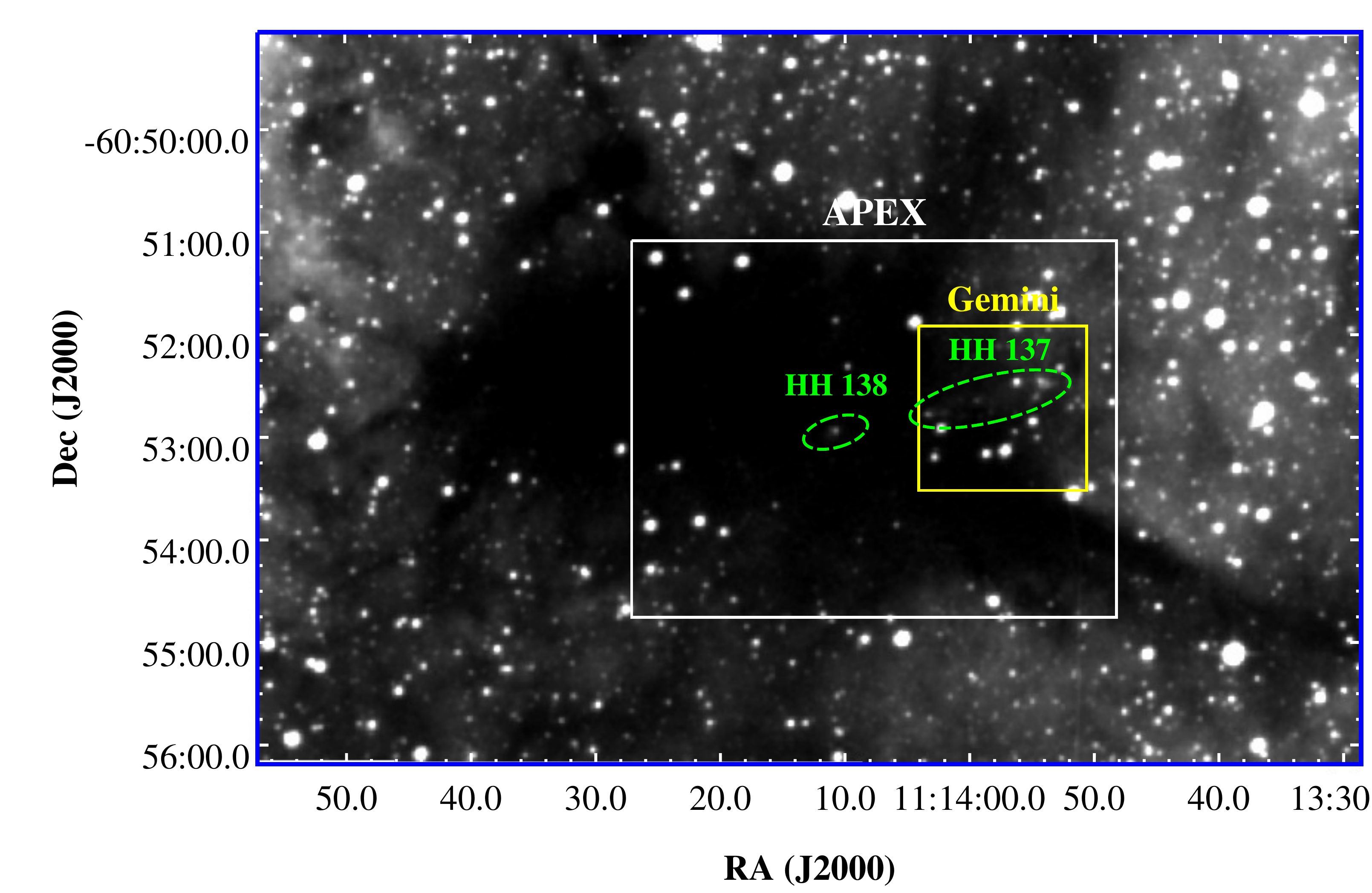}
    \caption{H$\alpha$ image from the SuperCOSMOS Survey, where the areas observed with the Gemini (yellow square) and APEX (white square) telescopes are marked. The position of the knots associated with HH\,137 and 138 are indicated  with ellipses (dotted green lines).}
    \label{fig_optico}
\end{figure}

Both HH objects are located towards the dark nebula \hbox{D291.4--0.2}\footnote{ D291.4--0.2 has also been catalogued as Globulet No.\,127 \citep{Sandqvist1977} and No.\,103 \citep{Feitzinger1984}, [DB2002b] G291.58+0.03 \citep{Dutra_Bica2002}, TGU~H1803 \citep{Dobashi2005} and DOBASHI 5797 \citep{Dobashi2011}.} \citep{Hartley1986}, in the Carina region. This highly dense dark cloud is $4\arcmin \times 2\arcmin$\ or $\sim2.5\times1.3$~pc (at a distance of 2.2~kpc, \citealt{Moffat1975,Steppe1977}) in size. It has been catalogued among the most opaque clouds studied by \cite{Sandqvist1977} as well as by \cite{Hartley1986} and also detected in a high density molecular tracer such as  \hbox{HCO$^+$\,J\,$=\,1\,\rightarrow\,0$} by \cite{Barnes2011}. In this latter work, this cloud is identified as BYF~129a.
Figure~\ref{fig_optico} shows the H${\alpha}$ image towards \hbox{D291.4--0.2} taken from the SuperCOSMOS Survey \citep{Parker2005}, where the dark cloud is observed
as an opaque patch with no background stars detected (in H${\alpha}$) due to the high extinction ($A_V = 13.5$, \citealt{Targon2011}). The green dotted ellipses mark the positions of HH\,137 and 138. The fields observed with the Gemini South and the APEX telescope are indicated with yellow and white boxes, respectively. The image reveals that HH\,137 and 138 coincide with the dark patch.

These HH objects, identified in [\ion{S}{ii}], H$\alpha$, and [\ion{N}{ii}] lines \citep{Ogura1993} show a chain of knots roughly in the East-West direction and they were named from A to J for HH\,137 and from A to D for HH\,138. \cite{Ogura1993} and \cite{Targon2011} have adopted a distance of 2.2~kpc \citep{Moffat1975,Steppe1977} for the two HH objects, and they estimated a size of 0.84 pc for HH\,137 and 0.23 pc for HH\,138.
Nevertheless, \cite{Jackson2008}, using the CS(2-1) high--density ($n\gtrsim10^{5}$~cm$^{-3}$) molecular line tracer, derived a kinematic distance of 1.37~kpc. Other authors have obtained similar distances  \citep[see e.g,][]{Kavars2005,Barnes2011,Planck_colaboration_2016}. Regarding the HH knots driving source, both \cite{Ogura1993} and \cite{Targon2011} suggested that it should be located between HH\,137-knot J and HH\,138-knot A. However, it is not clear whether HH\,138 and HH\,137 are linked to the same unknown driving source.

The aim of this paper is to detect the near-infrared counterparts of these HH objects as well as to search for likely molecular outflows associated with these objects and derive their physical parameters. Our new infrared and molecular data allow us to analyze the molecular environment linked to the HH objects and to contribute to a better understanding of the effect of outflows on the parent cloud.

The paper is outlined as follows. Our Gemini and APEX observations as well as \Spitzer, \WISE\ and \Herschel\ archive data used in this work are described in \textsection\,\,\ref{sec_data}. In \textsection\,\,\ref{sec_gemini}, we show  our  Gemini results for HH\,137. A complete analysis of HH\,137 and HH\,138 based on larger field mid- and far-IR archive images is presented in \textsection\,\,\ref{sec_infrared}. The search for the potential driving source of these HH objects is carried out in \textsection\,\,\ref{sec_driving_source}. The analysis of the molecular line emissions and the identification of outflows that are likely linked to HH\,137 and HH\,138 are described in  \textsection\,\,\ref{sec_apex}. In  \textsection\,\,\ref{sec_molecular_outflow} the physical parameters of the outflows as well as a scenario for the outflows configuration are presented.  Finally, in \textsection\,\,\ref{sec_summary} we summarize the work and highlight the main results.

%------------------------FIGURE-1---------------
%------------------------FIGURE-1---------------

%=========================================================================================
\section{Data sets} \label{sec_data}
% ----------------------------------------------------------------------------------------
\subsection{Gemini observations}

We used the NIR Gemini South Adaptive Optics Imager (GSAOI) and the Gemini Multi-conjugate Adaptive Optics System (GeMS), mounted at the 8-m Gemini South Telescope, in Cerro Pach\'on, Chile \citep{McGregor2004,Carrasco2012}. 
GeMS is an adaptive optic system that uses five sodium Laser Guide Stars (LSG), up to three Natural Guide Stars (NGS) and multiple deformable mirrors (DMs) that are optically conjugated with the main turbulence layers. This provides an adaptive optics (AO) corrected field that is larger than a single-conjugated adaptive optics (SCAO, \citealt{Neichel2014a,Rigaut2014}). 
GSAOI$+$GeMS provides diffraction limited images in the $0.9-2.4$\,\mum\ wavelength range over a $85\arcsec\times 85\arcsec$ field-of-view, with a imaging scale of $0\farcs0197$\,pixel$^{-1}$. The GSAOI detector is composed by $2\times2$ Rockwell HAWAII-2RG $2048\times2048$ pixel array mosaic with four gaps between the arrays of $\sim2.4$ mm, corresponding to $\sim2.5\arcsec$ on the sky. 

The images of HH\,137 were taken on 13th February 2014 (Program ID: GS-2014A-Q-29) in the H$_2$ (1-0 S(1), \hbox{$\lambda_{\text{c}} = 2.122$~\mum,} \hbox{$\Delta \lambda = 0.032$~\mum)} and K (\hbox{$\lambda_{\text{c}} = 2.200$~\mum,} \hbox{$\Delta \lambda = 0.340$~\mum)} filters. We used a $3\,\times\,3$ dither pattern with steps of $8''$ to remove the gaps between the detectors. We obtained a total of 9 science fields with individual exposure times of 100~sec each in the H$_2$ filter, and 9 science fields of 40~sec each in the  K-filter. 
Since it was possible to use three Natural Guide Stars (NGSs) with a good asterism geometry \citep{Neichel2014a,Neichel2014b}, the multi-conjugate adaptive optics (MCAO) performance was highly uniform over the field. The final images have a resolution of 0.09\,arcsec\,pixel$^{-1}$ which is in agreement with the values reported by \cite{Neichel2014a}.

The images were processed and combined using THELI\footnote{THELI is a tool for the automated reduction of astronomical images in optical, near- and mid-infrared, available on the website: \url{https://www.astro.uni-bonn.de/theli/}.} \citep{Schirmer2013,Erben2005}. 
The reduction process was similar to that described in \cite{Schirmer2013} and \cite{Schirmer2015}.
The science images were used to make a master sky image that, in turn, was subtracted.
To obtain the astrometry, the 2MASS catalog was used. We estimated an average precision of 0.17 arcsec in our coordinates with respect to the 2MASS catalog.

% ----------------------------------------------------------------------------------------
\subsection{APEX observations}

We obtained $^{12}$CO(3-2) (at 345.796~GHz), $^{13}$CO(3-2) (at 330.588~GHz), C$^{18}$O(3-2) (at 329.330~GHz), HCO$^+$(3-2) (at 267.557~GHz) and HCN(3-2) (at 265.886~GHz) molecular data using the 12-m Atacama Pathfinder EXperiment (APEX) telescope, located in the Llano de Chajnantor, Chile \citep{Gusten2006}. The observations were performed on 24th March 2016 and 17th to 20th June 2016 (Project 097.F-9707A-2016, PI: M. Rubio), using the \textit{On-The-Fly} technique with a space between dumps in the scanning direction of 9\arcsec. The off source position free of CO emission was \radec~=~(11:00:50.8, $-$61:00:19.4). The observed region was centered between HH\,137 and HH\,138 in the coordinate \radec~= ($11^h14^m07.38^s,-60\degr52\arcmin53.34\arcsec$) covering an area of $5'\times4'$ for $^{12}$CO, $^{13}$CO and C$^{18}$O, and $3'\times3'$ for HCO$^+$ and HCN. 

The APEX-2 (SHeFI, \citealt{Vassilev2008}) receiver was used to observe the CO lines with a half-power beam width (HPBW) of $\sim20''$, while the APEX-1 (SHeFI) receiver was employed to observe the HCO$^+$ and HCN line molecules with an HPBW of \hbox{$\sim24\arcsec$}.  

Raw data have a velocity resolution of $\sim 0.11$~\kms\ at 345~GHz. The velocity resolution was decreased to 0.3~\kms\ to improve the signal to noise. Pointing and calibration were performed during the observations using  L02 Pup, \hbox{IRAS\,07454--7112}, IRAS\,15194--5115, IRC\,$+$10216 and Carina sources. The calibration of the intensity has an uncertainty of 10\%.

%----------------------FIGURE-2-----------------------
\begin{figure*}
    \centering
    \includegraphics[width=\textwidth]{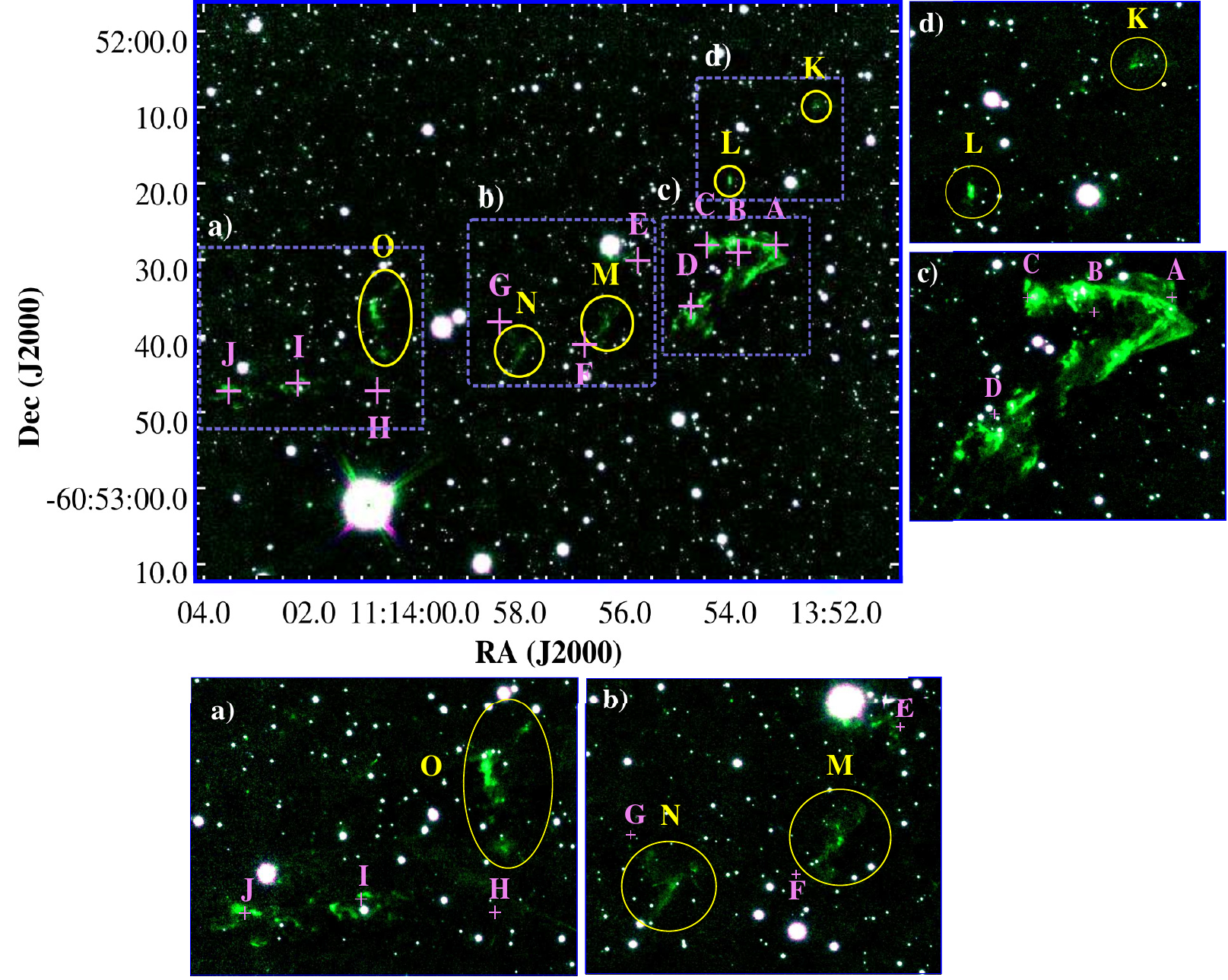}
    \caption{Composite image of HH\,137 showing K (magenta) and H$_2$ (green) filters taken with GSAOI+GeMS. The pink crosses mark the positions of the optical knots detected by \citet{Ogura1993}. The yellow circles and the ellipse indicate five new H$_2$ emission structures designated with the capital letters from K to O, belonging
    to MHO~1629. Enlarged sectors of the GSAOI+GeMS image, labeled as panels \textit{a)} to \textit{d)}, show more clearly the new H$_2$ knots, detected in this work as well as the H$_2$ counterparts of the HH\,137 optical knots. In particular panel \textit{c)} displays the terminal bow--shock shape, coinciding with HH\,137-A, B, C and D.}
    \label{fig_gsaoi}
\end{figure*}
%-----------------------------------------------------

The spectra were reduced using the Continuum and Line Analysis Single--dish Software (CLASS) of the Grenoble Image and Line Data Analysis Software (GILDAS\footnote{GILDAS is available at: \url{http://www.iram.fr/IRAMFR/GILDAS/}.}) according to the standard procedure of the CLASS software. The rms noise of the profiles after baseline subtraction and calibration is $0.2\,K$.
The observed line intensities are expressed as main-beam brightness temperatures $T_{mb}$ dividing the antenna temperature $T_{A}$ by the main-beam efficiency $\eta_{mb}$, equal to 0.72 for APEX--1 and APEX-2 \citep{Vassilev2008}.
The Astronomical Image Processing System (AIPS) package as well as the CLASS software were used to perform the analysis.

% ----------------------------------------------------------------------------------------
\subsection{Complementary data}

Images in the four IRAC bands centered at 3.6, 4.5, 5.8, and 8.0\,\mum\ were obtained from the Galactic Legacy Infrared Mid-Plane Survey Extraordinaire (GLIMPSE; \citealt{Benjamin2003}). Near-- and  mid--IR images at 3.4, 4.6, 12 and 22\,\mum\ were taken from the ALLWISE Data Release \citep{Wright2010}.  Far-IR data from the \Herschel\ Infrared GALactic Plane Survey \citep[Hi-GAL,][]{Molinari2010} at 70 and 160\,\mum\ \citep[PACS,][]{Poglitsch2010}, and at 250\,\mum\ \citep[SPIRE,][]{Griffin2010} were also used. These data have angular resolutions of 7\arcsec, 11\arcsec\ and 18\arcsec, respectively. The APEX Telescope LArge Survey of the GALaxy (ATLASGAL\footnote{ATLASGAL survey is available at: \url{http://atlasgal.mpifr-bonn.mpg.de/cgi-bin/ATLASGAL_DATABASE.cgi}.}, \citealt{Schuller2009}) at 870\,\mum\ (345 GHz) was also employed (beam size  19\farcs2). This survey has an rms noise in the range of 0.05-0.07 Jy\,beam$^{-1}$ and a calibration uncertainty of 15\%. In addition, optical images from SuperCOSMOS digitised data of the AAO/UKST H${\alpha}$ survey (SHS\footnote{SHS is available at: \url{http://www-wfau.roe.ac.uk/sss/halpha/}.}, \citealt{Parker2005}) of the Southern Galactic Plane were used to construct Figure~\ref{fig_optico}.

%==================================================================================================================
\section{Gemini results}    
\label{sec_gemini}

Figure~\ref{fig_gsaoi} shows a composite image of HH\,137 in K (magenta) and H$_2$ (green) in a $98''\times96''$ field. The chain of knots linked to HH\,137 is seen in shock excited H$_2$ emission. Enlarged sectors of Figure~\ref{fig_gsaoi} are displayed in panels \textit{a)} to \textit{d)}.  The pink crosses mark the position of optical knots listed by \cite{Ogura1993} with the nomenclature used therein. The new H$_2$ knots  are designated as MHO\,1629 in the on-line Catalogue of Molecular Hydrogen Emission-Line Objects\footnote{\url{http://astro.kent.ac.uk/~df/MHCat/}} \citep[MHOs,][]{Davis2010}. We assigned capital letters from K to O to the new structures, with no apparent optical counterparts. They are highlighted by yellow circles or an ellipse in Figure~\ref{fig_gsaoi}
 (see also \citealt{Ferrero2015b}). These new H$_2$ structures are more clearly seen in panels \textit{a)}, \textit{b)}  and \textit{d)}. Optical and NIR knots do not strictly coincide as it has been found in other jets \citep[e.g.,][]{Davis1994,Reipurth2000a}. Assuming that the exciting source is located between knot HH\,137-J and HH\,138-A \citep{Ogura1993}, the opening angle ($\theta$) of HH\,137 is $\sim13.2\degr$ and the collimation factor\footnote{The collimation factor $R_{\text{coll}}$ is defined as the ratio between the semi--major axis $R_{\text{max}}$ and the semi--minor axis $R_{\text{min}}$ of the outflow: \hbox{$R_{\text{coll}} = R_{\text{max}} / R_{\text{min}}$} \citep[see e.g.,][]{Lada1985}.} $R_{\text{coll}}\sim5$. We measure a PA of 284\degr\ in agreement with \citet[PA\,$\sim283\degr$]{Ogura1993}. 

Table~\ref{tab_gemini} provides a cross identification between the knots of \cite{Ogura1993} and the H$_2$ emissions in Figure~\ref{fig_gsaoi}. Considering that the width of the H$_2$ filter is about 10 times smaller than the width of the K-band filter, we calibrated in flux our H$_2$ (1-0) S(1) image using the 2MASS K$_s$ band magnitudes of 9 field stars. The fluxes were measured taking into account a 3$\sigma$ threshold between the background rms and H$_2$ emission. The knots coordinates and final H$_2$ fluxes are listed in fourth, fifth and sixth columns. The parameter $r$ (in the seventh column) is the size (radius in arcsec) of the circular area used for the aperture photometry of each knot. To estimate the errors, the uncertainties resulting from the conversion factor derived from the 2MASS K$_s$ magnitudes and the sky background variations in the H$_2$ filter were taken into account.

The high angular resolution obtained with GSAOI\,+\,GeMS/Gemini (see Figure~\ref{fig_gsaoi}) reveals the complex internal structure of the knots. The knots HH\,137-A to C display the typical terminal ``bow--shock'' shape more clearly than \citet[][ see panel \textit{c)}]{Ogura1993}. The panel \textit{a)} shows the HH\,137-H to J knots in more detail, whereas panel \textit{b)} displays HH\,137 knots E, F and G. We note that knots G and H are barely detectable in the Ogura's images. It is remarkable  that the spatial separation between knot H listed by \cite{Ogura1993} and the emission in H$_2$, identified as O, is of 10\farcs 6. For knot G and the nearest H$_2$ emission (N) a similar separation (of about 5\arcsec) is observed. These separations are not easily reconcilable with any velocity expected for these type of objects \citep[see e.g.,][]{Bachiller-Tafalla1999,Noriega-Crespo2001,Reipurth-Bally2001,Graham2003,McGroarty2007,Bally2016}.

%------------------------FIGURE-3---------------------
\begin{figure}
    \centering
    \includegraphics[width=\columnwidth]{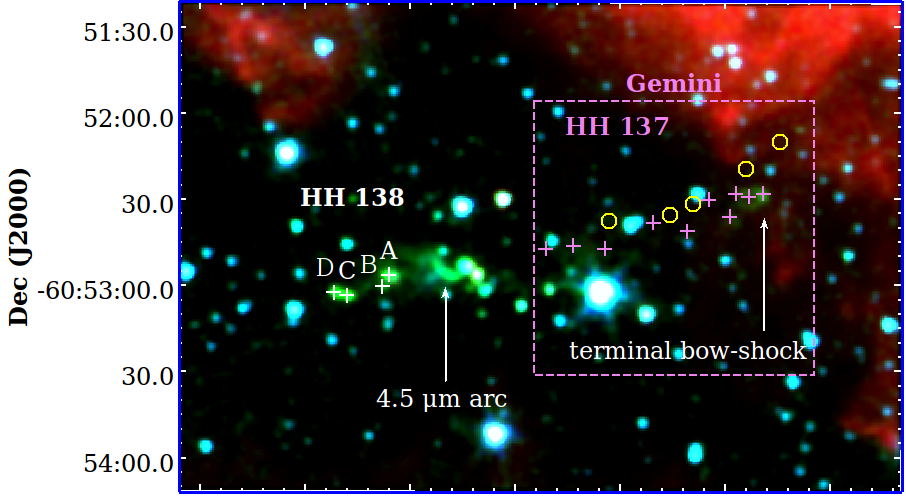}
    \includegraphics[width=\columnwidth]{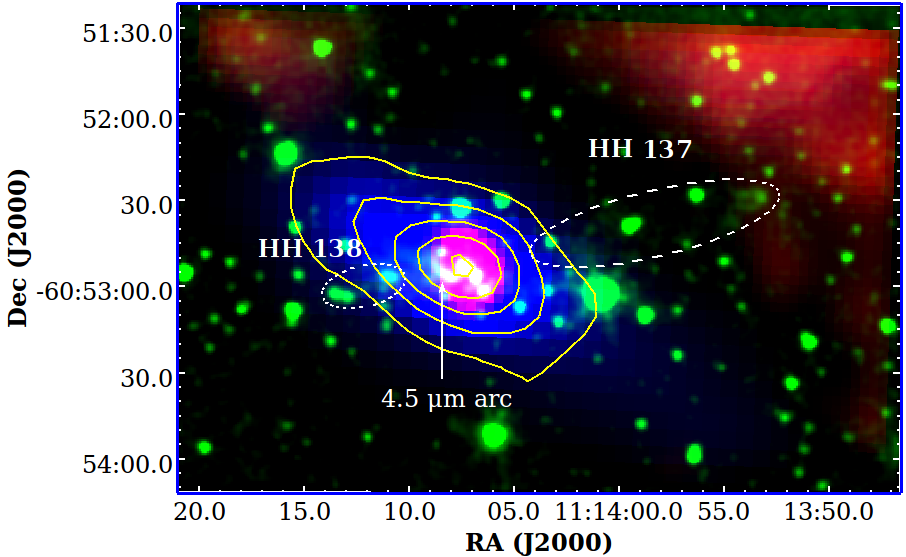}
    \caption{\textit{Upper panel:} Composite image of the \Spitzer\ emission at 3.6 (in blue), 4.5 (in green) and 8.0\,\mum\ (in red). The crosses mark the positions of Ogura\arcmin s knots and the circles the new H$_2$ knots found in this work (MHO~1629). The location of the arc-like structure detected in 4.5\,\mum\ and the location of H$_2$ terminal bow shock are indicated as well as the area covered with Gemini (dashed magenta line, see Figure~\ref{fig_gsaoi}).
    \textit{Lower panel:} \Spitzer\ 4.5~\mum, (in green), \Herschel\ 70 and 250~\mum\ (in red and blue, respectively), and ATLASGAL 870~\mum\ (in yellow contours). Contours correspond to 40, 50, 60, 70 and 80~Jy beam$^{-1}$. The ellipses (dashed white line) indicate the location of HH\,137 (on the right) and HH\,138 (on the left).}
    \label{fig_spitzer_egos}
\end{figure}
%------------------------FIGURE-----------------------
\begin{figure}
    \centering
    \includegraphics[width=\columnwidth]{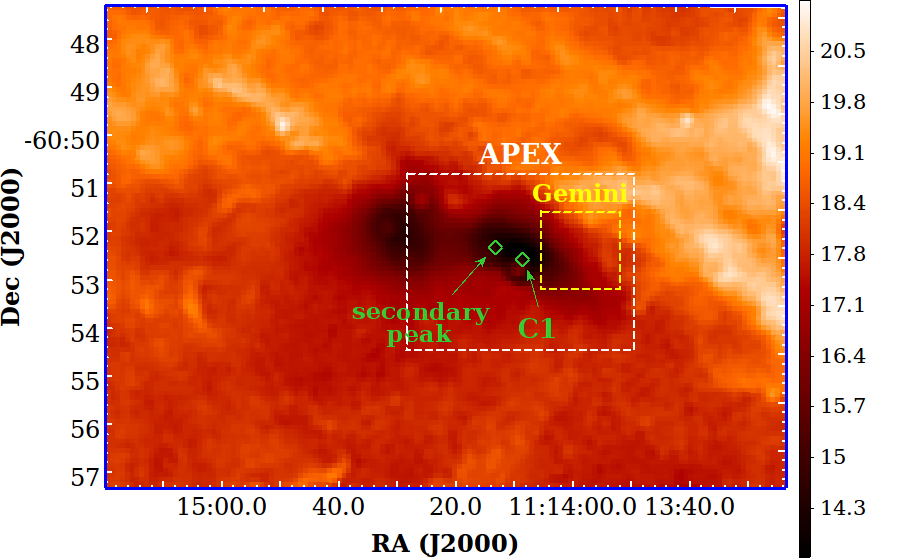}
    \includegraphics[width=\columnwidth]{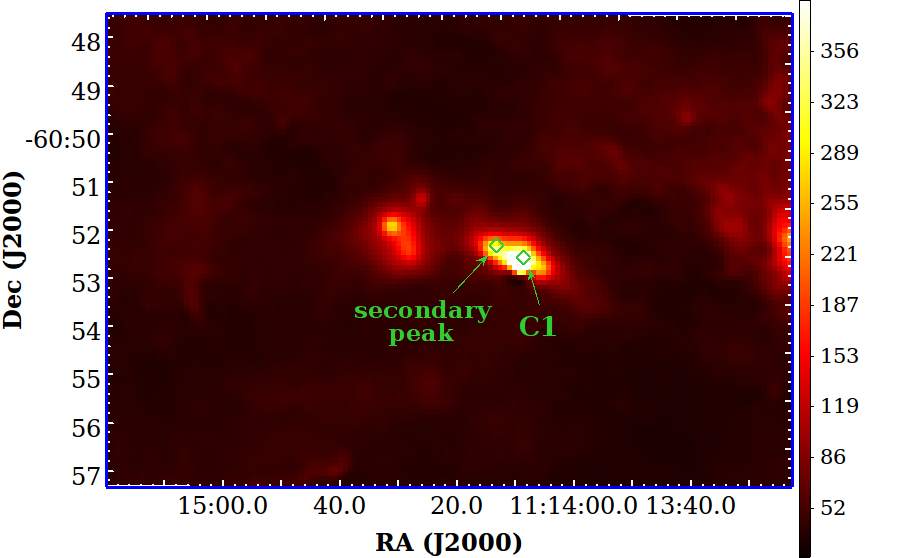}
    \caption{Dust temperature (upper panel) and column density (lower panel) maps in an area of $14\arcmin\times10\arcmin$ centered at \radec~$=$~(11:14:19.704, \hbox{--60:52:36.78)}, obtained from \protect\cite{Marsh2017}. Two structures are marked: C1, discussed in this section (Section \ref{sec_infrared}), and a secondary peak within the area of C1, introduced in Section \ref{sec_driving_source}. In the upper panel, the dashed yellow line box indicates the field observed with Gemini and the dashed white line box the field covered with APEX. Color scales are in units of K and $10^{20}$~cm$^{-2}$ for the temperature and column density maps, respectively.}
    \label{fig_temperature_density}
\end{figure}
%------------------------------TABL3-1----------------------
\begin{table*}
 	\centering
 	\caption[center]{HH\,137 knots detected by Ogura (1993) and in H$_2$ knots in our Gemini image.}
 	\label{tab_gemini}
 	\begin{tabular}{cccccccc}
 		\hline
 		  \multicolumn{3}{c}{Ogura (1993)} & \multicolumn{5}{c}{This work} \\
 		 \hline
 	HH\,137 & R.A.(J2000.0) & Dec.(J2000.0) & MHO\,1629 & R.A.(J2000.0) & Dec.(J2000.0) & \textit{F}$_{H_2}$ & r\\
 	      & ($^h:^m:^s$)  & ($\degr:\arcmin:\arcsec$) &      & ($^h:^m:^s$)  & ($\degr:\arcmin:\arcsec$) & (10$^{-5}$ Jy) & (\arcsec) \\
 		\hline
             A  & 11:13:53.14 & $-$60:52:28.08 & -- & -- & -- & -- & -- \\
            --  & -- & -- & A-1 & 11:13:53.42 & $-$60:52:30.25 & 28.8\,$\pm$\,5.2 & 4.09 \\
 	        --  & -- & -- & A-2 & 11:13:54.03 & $-$60:52:32.21 & 3.0\,$\pm$\,0.9 & 1.33 \\
 		     B  & 11:13:53.85 & $-$60:52:29.09 & B & 11:13:53.98 & $-$60:52:27.83 & 10.6\,$\pm$\,1.0 & 1.49 \\
 	    	 C  & 11:13:54.45 & $-$60:52:28.10 & C & 11:13:54.39 & $-$60:52:28.11 & 6.9\,$\pm$\,0.4 & 1.51 \\
 	    	 D  & 11:13:54.75 & $-$60:52:36.11 & D & 11:13:54.70 & $-$60:52:37.42 & 14.9\,$\pm$\,0.3 & 3.74 \\
 	    	 E  & 11:13:55.76 & $-$60:52:30.12 & E$^{\textit{a}}$ & 11:13:55.90 & $-$60:52:30.26 & -- & -- \\
	         F  & 11:13:56.77 & $-$60:52:41.14 & -- & -- & -- & -- & -- \\
 	         G  & 11:13:58.38 & $-$60:52:38.16 & -- & -- & -- & -- & -- \\
 	         H  & 11:14:00.70 & $-$60:52:47.20 & -- & -- & -- & -- & -- \\
 	    	 I  & 11:14:02.21 & $-$60:52:46.22 & I & 11:14:02.22 & $-$60:52:46.82 & 4.1\,$\pm$\,0.3 & 2.84 \\
 		     J  & 11:14:03.52 & $-$60:52:47.24 & J & 11:14:03.37 & $-$60:52:49.34 & 4.1\,$\pm$\,1.0 & 4.17 \\
 		    \hline
    	    --  & -- & -- & K$^{\textit{b}}$ & 11:13:52.43 & $-$60:52:10.12 & 1.0\,$\pm$\,0.1 & 1.20 \\
 		    --  & -- & -- & L$^{\textit{b}}$ & 11:13:54.03 & $-$60:52:19.45 & 1.8\,$\pm$\,0.1 & 1.02 \\
 		    --  & -- & -- & M & 11:13:56.51 & $-$60:52:39.80 & 2.1\,$\pm$\,0.5 & 3.11 \\
 		    --  & -- & -- & N & 11:13:58.06 & $-$60:52:42.15 & 2.3\,$\pm$\,0.7 & 3.02 \\
 		    --  & -- & -- & O & 11:14:00.73 & $-$60:52:37.03 & 6.7\,$\pm$\,0.3 & 2.61 \\
 		\hline
 	\end{tabular}
 	\vskip 0.05 cm
\noindent {\footnotesize{Notes: $^{\textit{a}}$ Knot E was not measured as the H$_2$ emission is contaminated by a close bright star.\\
$^{\textit{b}}$ Knots K and L correspond to knots X2 and X1 in \cite{Ferrero2015b}.}}
\end{table*}
%----------------------------------------------------------
%------------------------------TABLE-2----------------------
\begin{table*}
    \centering
    \caption{HH\,138 knots detected by Ogura (1993) and new \emph{EGOs} at 4.5\,\mum.}
 	\label{tab_EGOS_HH138}
 	\begin{tabular}{rcccccc}
	\hline
 		 & \multicolumn{2}{c}{Ogura (1993)} & \multicolumn{4}{c}{This work}  \\
 		 \hline
 	Knot & R.A.(J2000.0) & Dec.(J2000.0) & R.A.(J2000.0) & Dec.(J2000.0) & \textit{F}$_{[4.5]}$ & Area\\
 	     & ($^h:^m:^s$)  & ($\degr:\arcmin:\arcsec$) & ($^h:^m:^s$)  & ($\degr:\arcmin:\arcsec$) & (mJy) & (arcsec$^2$) \\
 		\hline
 	 HH\,138-A  & 11:14:10.99 & $-$60:52:56.35 & 11:14:11.00 & $-$60:52:56.76 & 479.7\,$\pm$\,20.5 & 163.63 \\
   		    B  & 11:14:11.29 & $-$60:53:00.35 & 11:14:11.25 & $-$60:53:00.41 & 54.6\,$\pm$\,4.1 & 33.55 \\
 	    	C  & 11:14:13.00 & $-$60:53:03.38 & 11:14:12.92 & $-$60:53:03.67 & 274.9\,$\pm$\,1.6 & 72.62 \\
 		    D  & 11:14:13.61 & $-$60:53:02.39 & 11:14:13.57 & $-$60:53:02.65 & 369.2\,$\pm$\,1.2 & 91.18 \\
 		\hline
 	4.5--arc   & -- & -- & 11:14:08.42 & $-$60:52:55.36 & 277.8\,$\pm$\,82.6 & 296.73 \\
 	    \hline
 	HH\,137-A-B-C & -- & -- & 11:13:53.70 & $-$60:52:29.47 & 374.4\,$\pm$\,19.0 & 452.74 \\
 		\hline
    \end{tabular}
\end{table*}

%=========================================================================================
\section{IR data analysis}  \label{sec_infrared}

The upper panel of Figure~\ref{fig_spitzer_egos} displays a composite image of the \Spitzer\ emission at 3.6 (in blue), 4.5 (in green) and 8.0\,\mum\ (in red). Optical knots associated with HH\,137 and 138 from \cite{Ogura1993} are marked with crosses (magenta and white crosses, respectively), while the yellow circles indicate the new H$_2$ knots detected (see Figure~\ref{fig_gsaoi}). Several bright 4.5\,\micron\ emissions are evident, most of them coinciding with Ogura's knots and/or the H$_2$ knots here reported associated with HH\,137. The optical knots HH\,137-A to D belonging to the H$_2$ bow--shock region are the most prominent at 4.5\,\micron. An arclike structure located between HH~137 and HH~138 is clearly seen at 4.5~\mum. This structure has not been previously reported. Thus, we note it for the first time and name it as the 4.5~\mum\ arc. 
The projected size of the arc is of about 12\arcsec\ (0.09~pc at 1.5~kpc\footnote{Hereafter we will adopt a distance of 1.5~kpc for HH\,137 and 138, as discussed in Section \ref{sec_co_data}.}, see Section \ref{sec_co_data}), with the apex pointing roughly to the west. 

Bright extended emissions in 4.5\,\mum\ are usually classified as \emph{EGOs} (Extended Green Objects, \citealt{Cyganowski2008}) or \emph{``Green Fuzzies''} \citep{Chambers2009}. Since the 4.5\,\micron\ band contains several rotovibrational H$_2$ lines and CO band heads \citep{Reach2006,Smith-Rosen2005,Watson2010}, it is generally used as shock diagnostic. The fact that several 4.5\,\mum\ emissions in Figure~\ref{fig_spitzer_egos} spatially coincide with the H$_2$ emissions in HH\,137 (MHO 1629), and particularly with the HH\,137 terminal bow--shock region, indicates collisional excited objects.  Table~\ref{tab_EGOS_HH138} lists the HH\,138 knots identified by \cite{Ogura1993} and the new \emph{EGOs} shown in Figure~\ref{fig_spitzer_egos}, upper panel. Consequently, even when our H$_2$ image does not cover the HH\,138 region, the coincidence between 2.12 and 4.5\,\mum\ emissions suggests that the 4.5\,\mum\ emission is associated with jet emission \citep[e.g.,][]{Smith2006,Davis2007,Cyganowski2008}.

The emission at 8\,\mum\ traces polycyclic aromatic hydrocarbons (PAHs) and photodissociation regions (PDRs, \citealt{Watson2008}). Emission at 8\,\mum\ is identified towards the borders of the image and in coincidence with part of the terminal shock and some point-like sources close to the 4.5\,\mum\ and optical knots of HH\,138. The lack of extended 8\,\mum\ emission close to the HH objects indicates that the region has a large visual absorption and shares some characteristics with infrared dark clouds \citep[IRDC,][]{Sanhueza2010}. In fact, the GLIMPSE [3.6] and [4.5] $\mu$m  mosaic image of the region clearly delineates a dark patch, coinciding with the optical extinction. Moreover, \cite{Simon2006a} had previously reported the detection of an IRDC named MSX-DC G291.40-0.19 based on 8.3\,\micron\ mid--infrared image acquired with the Midcourse Space Experiment satellite.

The composite image in the bottom panel of Figure~\ref{fig_spitzer_egos} displays \Spitzer\ 4.5\,\mum\ (in green), \Herschel\ 70 and 250\,\mum\ (in red and blue, respectively), and ATLASGAL 870\,\mum\ (in contours) data. The dashed white ellipses mark the positions of HH\,137 and HH\,138. A bright isolated source is clearly detected at 70\,\mum, which coincides with the two point-like sources close to the 4.5\,\mum\ arc (see the upper panel). The emission distribution from ATLASGAL reveals a cold dust clump (G291.367--00.214, \citealt{Csengeri2014}) centered on the arc at 4.5\,\mum\  and a group of \Spitzer\ point-like sources. The dust clump is $100\arcsec \times 46\arcsec$ or about 0.73\,pc\,$\times$\,0.33\,pc at 1.5~kpc). This clump is classified as a protostellar region since it contains several IR sources detected at 4.5\,\mum\ \citep{Chambers2009}. 

Using ATLASGAL 870~\mum\ data, the dust mass of the clump can be estimated applying the expression by \cite{Hildebrand1983}:
\begin{equation}
\text{M}_{\text{dust}} = \frac{S_{870} \ d^{2} }{\kappa_{870} \ B_{870}(T_{\text{dust}})} ,
\end{equation}
\noindent where $S_{870}$ is the flux density at 870\,\mum, $d = 1.5\pm0.5$~kpc, $\kappa_{870}~=~2.06$~cm$^{2}$/g is the dust opacity per unit mass \citep{Ossenkopf1994} and $B_{870}(T_{\text{dust}})$ is the Planck function at a temperature $T_{\text{dust}}$. 
The flux density, $S_{870}$, of the clump was obtained integrating the observed emitting area within an aperture of $45\arcsec$ (radius), resulting in a flux of $6.0\,\pm\,0.5$~Jy. The subtracted background emission was estimated in an annulus with inner and outer radii of 90\arcsec and 140\arcsec, respectively.

To estimate $T_{\text{dust}}$ of the clump, we used temperature as well as column density maps from \cite{Marsh2017}. These maps\footnote{\url{http://www.astro.cardiff.ac.uk/research/ViaLactea/}} were produced employing the point process mapping (PPMAP) method based on a Bayesian procedure applied to Herschel continuum data in the wavelength range $70-500$~\mum\ \citep{Marsh2015}.
This method provides resolution--enhanced ($\sim12\arcsec$) images of dust temperature and column density. Figure~\ref{fig_temperature_density} shows both the temperature and column density (N(H$_2$)) maps, in an area of $14\arcmin \times 10\arcmin$ centered at \hbox{\radec~$=$~(11:14:19.704, --60:52:36.78)}. In both panels of Figure~\ref{fig_temperature_density} the dust clump C1 is marked. A secondary weaker dust peak is detected within the area of C1.

The temperature within the area of the dust clump varies from 13.5 to 16.2 K. Adopting an average temperature of \hbox{$14.8 \pm 0.6$~K}, the dust mass turns out to be \hbox{$M_{dust}~=~1.2\,\pm\,0.5$~\msun}. For a standard gas-to-dust ratio, $R$ of $100$ \citep[e.g.,][]{Elia2013, Elia2017, Konig2017}, the hydrogen gas mass amounts to \hbox{$M_{H_2}=120\,\pm\,50$~\msun}.  It is worth  mentioning that the quoted uncertainty in the hydrogen mass neglects any known factors in $R$ and/or $\kappa_{870}$.
We also used the dust column density map from \cite{Marsh2017}, shown in Figure~\ref{fig_temperature_density} lower panel, to estimate $M_{H_2}$. To do this, we integrated over the area delimited by the contour of $8.7\times 10^{21}$ cm$^{-2}$ at $3\sigma$. This contour comprises an area similar to that used to calculate the flux at 870~\mum. The hydrogen gas mass turns out to be $M_{H_2}=175\,\pm\,21$~\msun. 

The discrepancy between the two values of mass is larger than the quoted uncertainties.
Nonetheless, other authors have also found differences between the masses derived from the APEX 870~\mum\ flux and the Herschel H$_2$ column density, in the sense that 870~\mum\ masses are lower than H$_2$ column density masses \citep{Deharveng2010,Liu2017,Dewangan2017,Das2018}. In particular, \cite{Liu2017} attributed this discrepancy to possible drawbacks in ground-based data reduction that underestimate the mass, due to the loss of large-scale emission when applying the sky noise subtraction. Hence, the mass derived from the 870~\mum\ flux may only represent a lower limit to the true mass. Moreover, the PPMAP procedure provides differential column density as a function of dust temperature and position  (see \citealt{Marsh2015}). Consequently, hydrogen mass derived by the PPMAP method is probable more reliable and accurate than that calculated from the 870~\mum\ ATLASGAL flux density.

%-------------------------------------------------------------------------------------
\section{The driving source}  \label{sec_driving_source}

\cite{Ogura1993} proposed that the driving source of HH\,137/138 should be located between HH\,137-knot J and HH\,138-knot A. The distance between the knots is about $1\arcmin$. Bearing this in mind, we searched for all the young stellar objects (YSOs) that are likely to be the driving sources in an area of \hbox{$60\arcsec \times40\arcsec$}. A number of YSOs were identified using the \Spitzer\ and \WISE\ point source catalogs. The \Spitzer\ candidates were obtained from the MYStIX IR Excess Source catalog  from \cite{Povich2013}, who identified a few YSOs in the selected region. We have placed these YSOs in a Color-Color ([3.6]--[4.5] vs \hbox{[5.8]--[8.0]})  diagram  and used the criteria by \cite{Allen2004} to determine their evolutionary status. The \WISE\ candidates, in the same area, were obtained from \cite{Cutri2013} applying the criteria by \cite{Koenig2012}. Class I candidates are protostars surrounded by dusty infalling envelopes, while Class II candidates are dominated by accretion disks. Both Class I as well as Class II young stars have been identified as the exciting sources of optical jets and molecular outflows \citep{Bachiller-Tafalla1999,Reipurth-Bally2001,Arce2007}.

\begin{figure}
    \centering
    \includegraphics[width=\columnwidth]{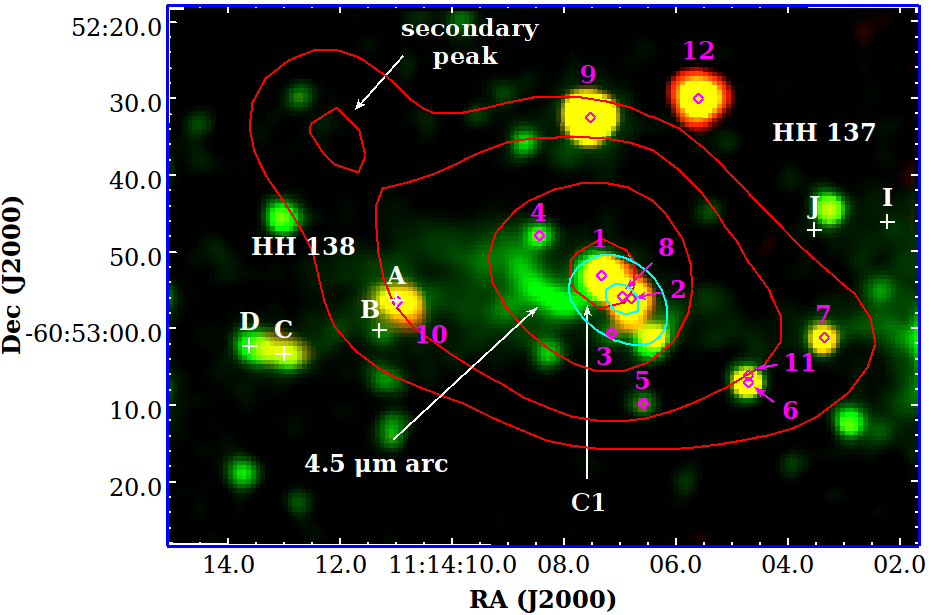}
    \caption{Overlay of the \Spitzer\ 4.5 (in green) and 8\,\mum\ (in red) images, and contours of \Herschel\ 160\,\mum\ (in red) and \WISE\ 4.6\,\mum\ (in cyan) images. The contours correspond to 0.05, 0.1, 0.3 and 0.8 Jy pixel$^{-1}$ for \Herschel, and 0.05 and 0.07 Jy pixel$^{-1}$ for \WISE. The peak at 160\,\mum\ is labeled as C1 at \hbox{\radec~=~(11:14:07.324, --60:52:52.90)}. A secondary weaker peak is also shown to the north--east at  \hbox{\radec~=~(11:14:12.054, --60:52:35.50)}. The pink diamonds and numbers correspond to the sources in Table~\ref{tab_ysos}. The white crosses mark the optical knots of HH\,137 and HH\,138 from \citet{Ogura1993} and the white arrows indicates the 4.5~\mum\ arc, the dust clump C1 and the secondary peak.}
    \label{fig_driving_sources}
\end{figure}

\Spitzer\ and \WISE\ sources are labeled in Fig. \ref{fig_driving_sources}, which is an overlay of the 4.5 (in green) and 8\,\mum\ (in blue) \Spitzer\ images, 160\,\mum\ (in red contours) \Herschel\ image and 4.6\,\mum\ (in cyan contours) \WISE\ image. Coordinates, fluxes in the four bands of each catalog, evolutionary status and cross-correlations of the \Spitzer\ and \WISE\ sources are listed in Table~\ref{tab_ysos}. All the \WISE\ sources are also detected in the \Spitzer\ catalog. The slight differences in position are attributed to the angular resolution of the corresponding instruments. \Spitzer\ sources 1 and 2 (or \WISE\ source 8) coincide with the peak of emission in \Herschel--PACS 160\,\mum\, identified as C1 (see Section \ref{sec_infrared}) that also clearly shows up in 70\,\mum~ (see Figure~\ref{fig_spitzer_egos}, lower panel). The coincidence of the emissions at 160 and 70\,\mum\ with two YSO candidates suggests that dust temperature in this region is higher than in the rest of the dust clump, giving additional support to the YSO status of these sources. In addition, \cite{Walsh2014} found H$_{2}$O maser emission coincident with \Spitzer\ source 1 (see Table~\ref{tab_ysos}), thus reinforcing its YSO classification.

\WISE\ source 8, which is likely to be associated with \emph{EGOs}, is bright at 22\,\mum\ (3.08 mag, see Table~\ref{tab_ysos}) and lies within the dark infrared cloud (see Section \ref{sec_infrared}). According to the classification scheme of \cite{Chambers2009} it would be considered as an \emph{``active'' core}. \WISE\ data at 22\,\mum\ are being used instead of at 24\,\mum, as was originally proposed by \cite{Chambers2009}, since \Spitzer\ MIPS images at 24 $\mu$m at the source position are not available. 

The projected separation between \Spitzer\ sources 1 and 2 is 5\arcsec, or $\sim$\,7500 AU (at 1.5~kpc). Young low mass binaries have separations from 500 to 4500~AU (see \citealt{Connelley2008} and references therein). Young massive binaries should expand a larger range of separations since the components are larger. Recently, \cite{Pomohaci2019} performed a pilot survey of candidate wide young massive binary stars, finding that pairs (with a probability of chance alignment $<$ 20\%) have separations between $\sim 1300$ and $\sim 13000$~UA. Thus, \Spitzer\ sources 1 and 2 might be gravitationally bound, forming a wide binary system, assuming they are massive stars.

In summary, we propose that  \Spitzer\ sources 1 and 2, coinciding with \WISE\ source 8 (WISE\,J111406.96--605255.9), are the candidate powering sources of the HH objects. \Spitzer\ sources 1 and 2 are also very close to the 4.5\,\mum\ arc-like structure and are projected onto the axis of HH\,137. \WISE\ source 8 has been also classified as YSO candidate  in the AKARI/FIS young stellar objects catalog \citep{Toth2014}.

%--------------------FIGURE-4---------------------
% -------------------------------------------------------------------
%------------------------------TABLE-3----------------------- 
\begin{table*}
\begin{center}
\caption{YSO candidates projected onto the central region of HH\,137/138.} \label{tab_ysos}
\begin{tabular}{cccccccccc}
\hline
\multicolumn{10}{c}{\Spitzer\ candidates} \\
\hline
$\#$ &  R.A.    & Dec.   &   \Spitzer &  \multicolumn{4}{c}{Fluxes [mag]} & YSO Class & Ref. \WISE\ \\
     & ($^h$:$^m$:$^s$) & (\degr:\arcmin:\arcsec) & \textit{name} &  $[3.6]$ & $[4.5]$ & $[5.8]$ & $[8.0]$ &  & $\#$ \\
\hline
1  & 11:14:07.33 & $-$60:52:53.25 & G291.3671$-$00.2137 & 10.569 &  9.877 &  9.300 &  8.599 & I & 8 \\
2  & 11:14:06.80 & $-$60:52:56.12 & G291.3664$-$00.2148 & 12.533 & 11.111 & 10.027 &  9.155 & I & 8 \\
3  & 11:14:07.15 & $-$60:53:00.75 & G291.3671$-$00.2157 & 14.760 & 13.516 & 12.933 &  11.895 & I & -- \\
4  & 11:14:08.44 & $-$60:52:47.97 & G291.3687$-$00.2115 & 13.092 & 12.497 & 11.986 & 11.349 & II & -- \\
5  & 11:14:06.58 & $-$60:53:09.82 & G291.3674$-$00.2185 & 14.004 & 13.088 & 12.500 & 11.592 & I & -- \\
6  & 11:14:04.70 & $-$60:53:07.05 & G291.3635$-$00.2192 & 12.396 & 11.604 & 10.922 & 10.174 & II & 11\\
7  & 11:14:03.35 & $-$60:53:01.31 & G291.3604$-$00.2187 & 12.874 & 11.931 & 11.154 & 10.411 & II & -- \\
\hline
\multicolumn{10}{c}{\WISE\ candidates} \\
\hline
$\#$ & R.A.    & Dec.    &   {\WISE } &   \multicolumn{4}{c}{Fluxes [mag]} & YSO Class & Ref. \Spitzer\ \\
 & ($^h$:$^m$:$^s$) & (\degr:\arcmin:\arcsec) & \textit{name}&  $[3.4]$ & $[4.6]$ & $[12.0]$ & $[22.0]$ &  & $\#$ \\
\hline
8  & 11:14:06.96 & $-$60:52:55.96 & J111406.96$-$605255.9 & 12.176 &  9.653 & 8.106 & 3.08 & II & 1-2\\
9  & 11:14:07.53 & $-$60:52:32.57 & J111407.53$-$605232.5 & 9,849 &  9.155 & 9.78 & 6.882 & II & -- \\
10 & 11:14:10.98 & $-$60:52:56.53 & J111410.98$-$605256.5 & 13.065 & 10.973 & 10.278 & 6.242 &  II & -- \\
11 & 11:14:04.70 & $-$60:53:06.25 & J111404.69$-$605306.2 & 13.144 & 11.365 & 10.274 & 6.506 &  II & 6\\
12 & 11:14:05.59 & $-$60:52:30.08 & J111405.58$-$605230.0 & 11.759 & 10.589 & 7.90 & 5.081 &  I & -- \\
\hline
\end{tabular}
\end{center}
\end{table*}
%------------------------------TABLE-3------------------- 

%=========================================================================================
\section{APEX molecular gas results}  \label{sec_apex}
%--------------------------- 

\subsection{Analysis of the CO data} \label{sec_co_data}

\begin{figure*}
    \centering
    \includegraphics[width=14cm]{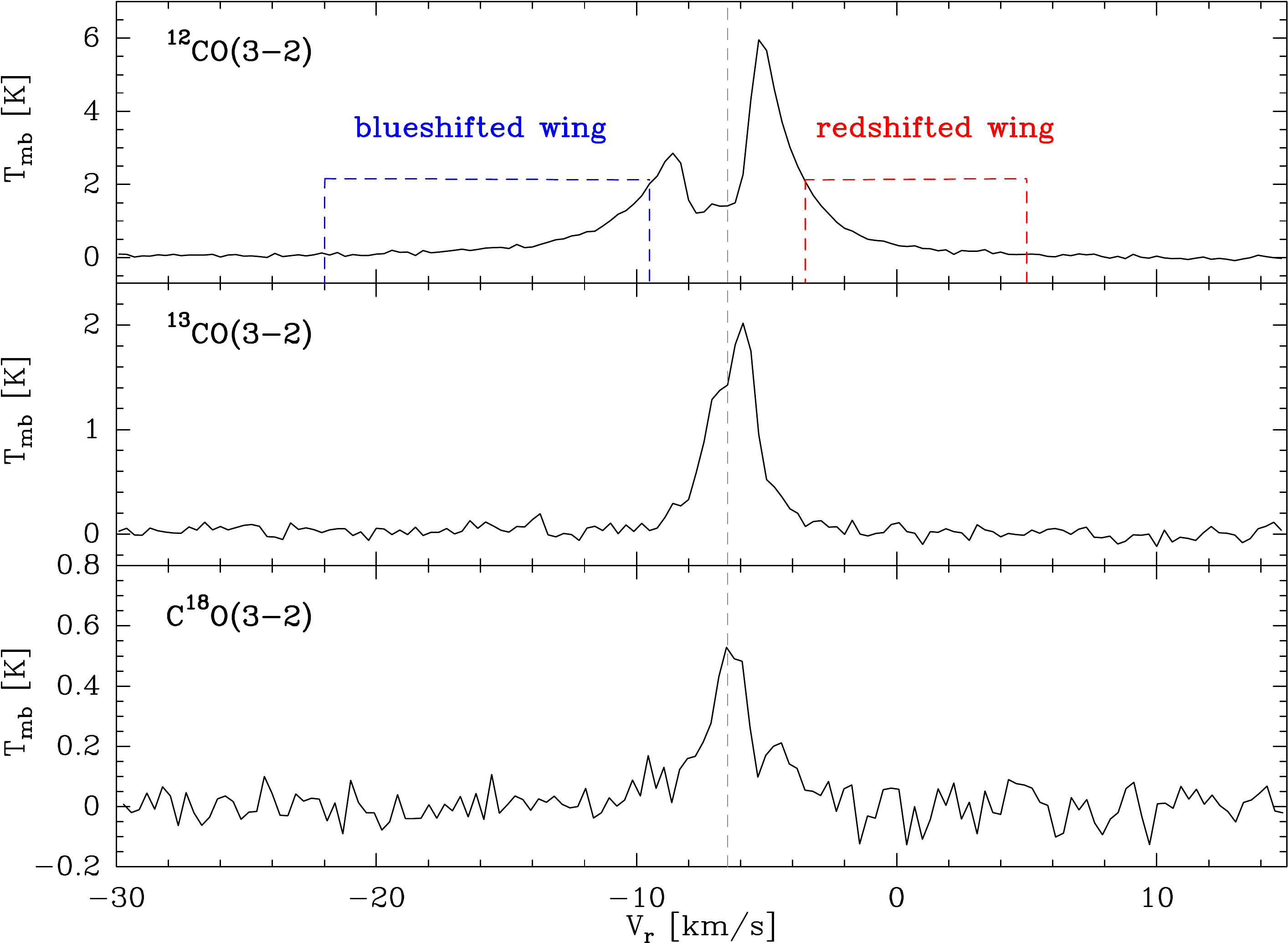}
    \caption{Averaged profiles of $^{12}$CO(3-2), $^{13}$CO(3-2) and C$^{18}$O(3-2) within the emitting region in the velocity interval [--30,+15]~\kms\ showing main-beam brightness temperature T$_{mb}$ vs. LSR velocity. The vertical dashed gray line indicates the velocity position corresponding to the emission peak of C$^{18}$O(3-2) taken as the systemic velocity. The dashed blue and red line boxes in the upper panel show the $^{12}$CO(3-2) wings of molecular outflows (see Section \ref{sec_molecular_outflow}).}
    \label{fig_perfil_co}
\end{figure*}

The $^{12}$CO(3$-$2) line shows emission above $3\sigma$ between \hbox{$\approx -22$} and $+5$~\kms\ (all velocities are referred to the LSR).  The averaged spectra of the three CO isotopologues  ($^{12}$CO, $^{13}$CO and C$^{18}$O), taken in an aperture of $132''\times81''$ centered at \hbox{\radec\,$=$\,(11:14:06.497,} \hbox{--60:52:40.70)} and covering most of the CO emission, are shown in Figure~\ref{fig_perfil_co}. The $^{12}$CO profile shows two peaks at --8.6 and --5.2~\kms, with a depression at --7.1~\kms.  The blueshifted and redshifted wings are indicated with dashed line boxes and are associated with molecular outflows (see Section \ref{sec_molecular_outflow}).

The $^{13}$CO(3$-$2) line shows a shallower depression than the $^{12}$CO(3$-$2) line profile centered at a similar velocity. The \hbox{C$^{18}$O(3-2)}\footnote{C$^{18}$O is a high-density molecular tracer with a critical density of \hbox{$\sim1.2\times10^{4}$~cm$^{-3}$} in J$~=~3-2$ estimated from \cite{Shirley2015}, assuming a kinetic temperature of $\sim$ 15 K.} is centered at \hbox{--6.3}~\kms, redshifted with respect to the $^{12}$CO depression. As expected, the $^{13}$CO and C$^{18}$O profiles are narrower (in velocity) and weaker (in T$_{mb}$) than the $^{12}$CO profile and have emissions within the velocity intervals \hbox{[--9.5, --3.6]~\kms}, and \hbox{[--8,--4]~\kms}, respectively.

Taking into account the Galactic non--circular motions \citep{Brand1993} and a systemic velocity of \hbox{--6.3}~\kms\ (obtained from the C$^{18}$O peak) with an uncertainty of 5~\kms, we estimated the kinematic distance. As the Carina region is located in the fourth quadrant at the inner Galaxy, there are two kinematic distances along the line of sight: the ``near'' and ``far''.  Since the region has a high extinction (see Figure~\ref{fig_optico}) and has been catalogued as an IRDC \citep{Simon2006b,Jackson2008,Dobashi2011}, HH~137 and HH~138 should lie in front of the Galaxy and near the solar circle. Then, we adopt the ``near'' kinematic distance that turns out to be equal to 1.5\,$\pm$\,0.5\,kpc. This value is compatible with those calculated by \cite{Kavars2005,Jackson2008,Barnes2011,Planck_colaboration_2016} and roughly agrees with the distance adopted by \citet{Ogura1993} of $2.2\,\pm\,0.2$~kpc.

\begin{figure*}
    \centering
    \includegraphics[width=0.33\textwidth]{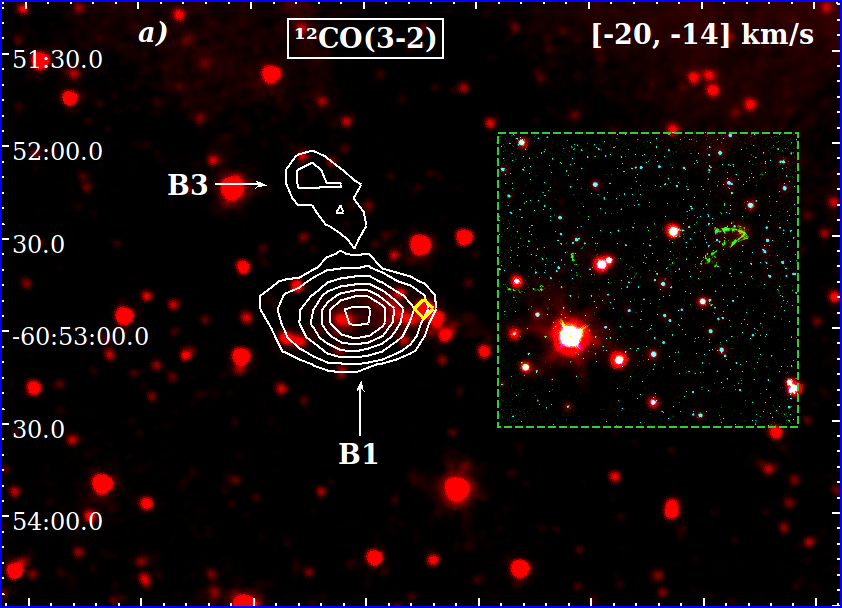}
    \includegraphics[width=0.33\textwidth]{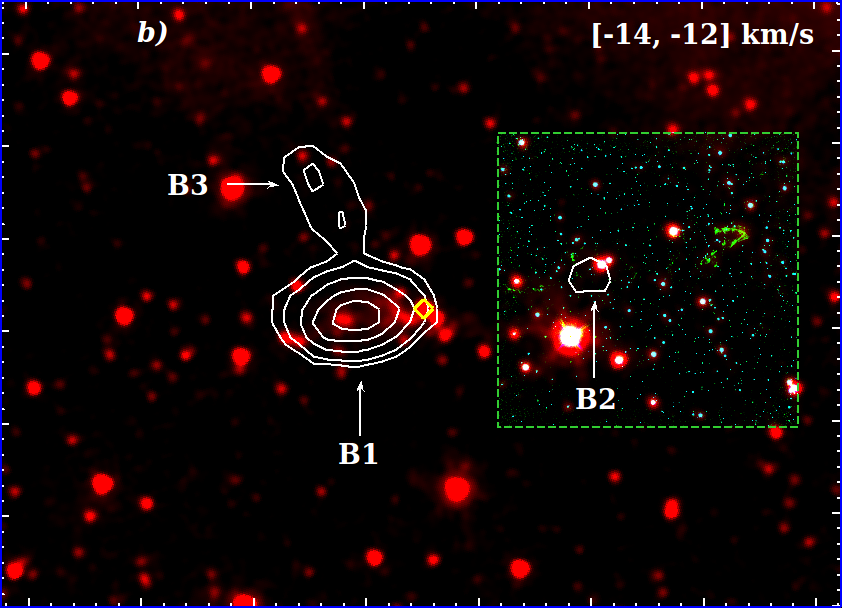}
    \includegraphics[width=0.33\textwidth]{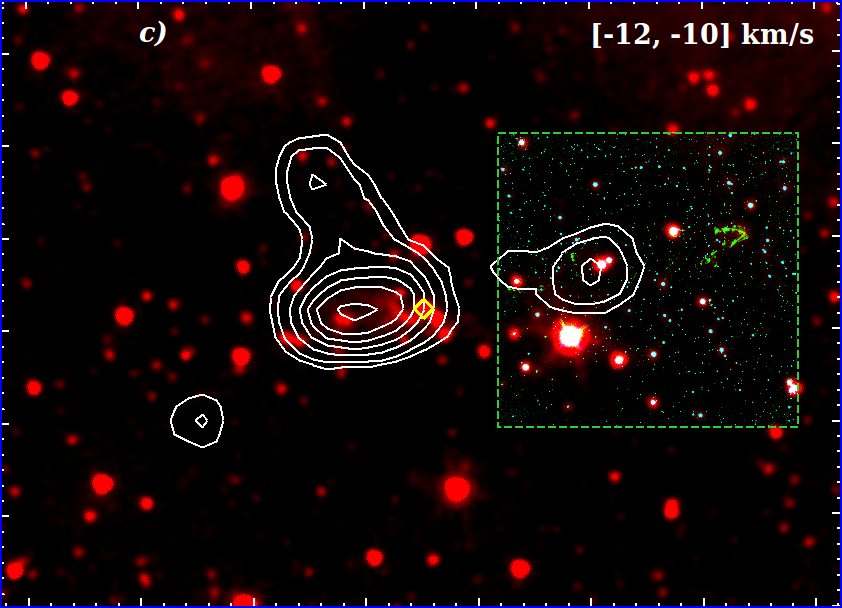}\\
    \includegraphics[width=0.33\textwidth]{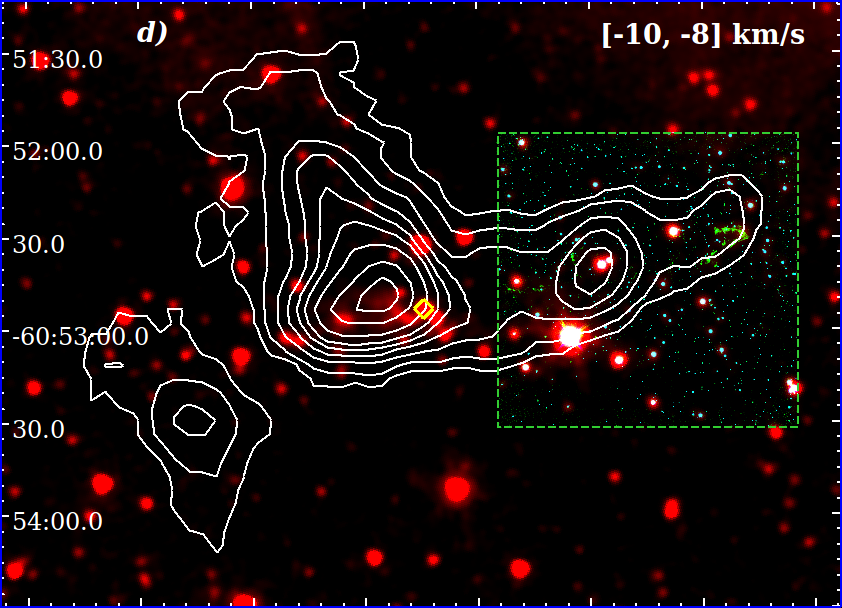}
    \includegraphics[width=0.33\textwidth]{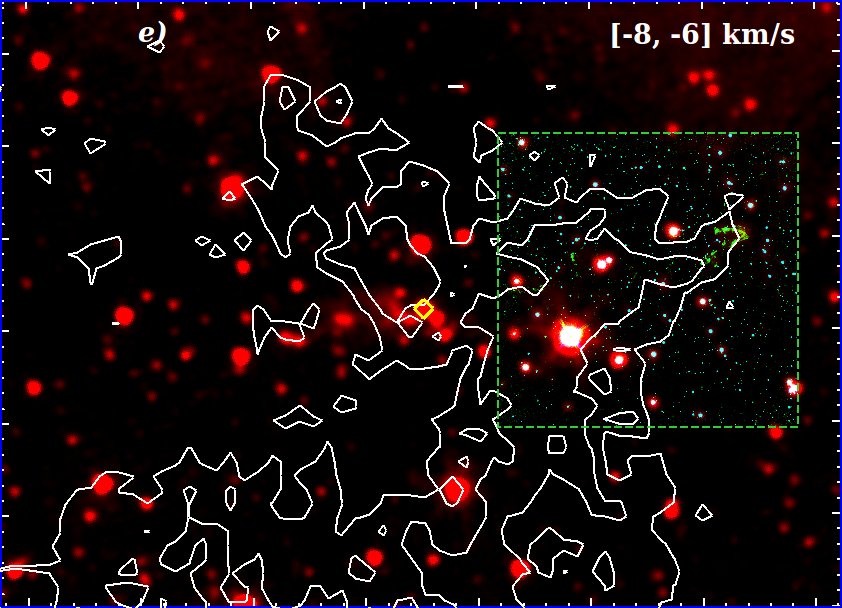}
    \includegraphics[width=0.33\textwidth]{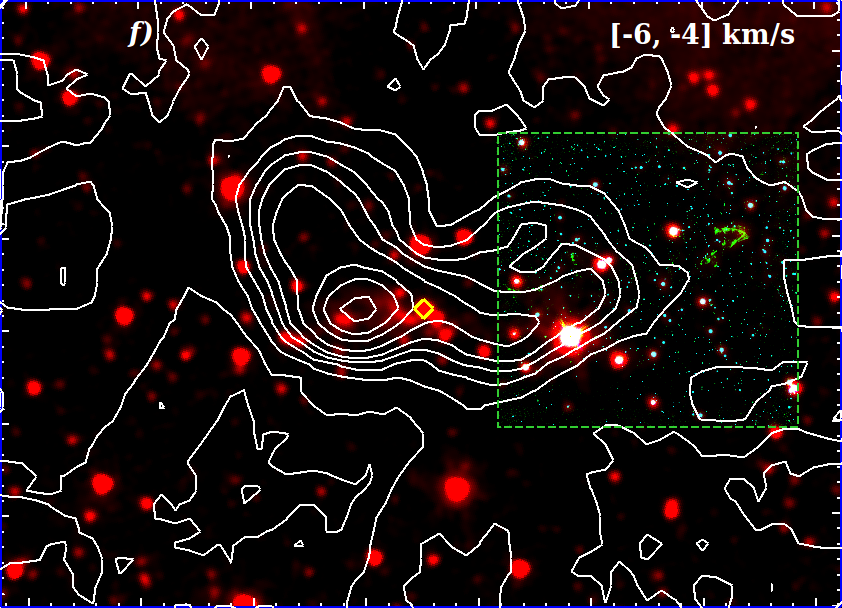}\\
    \includegraphics[width=0.33\textwidth]{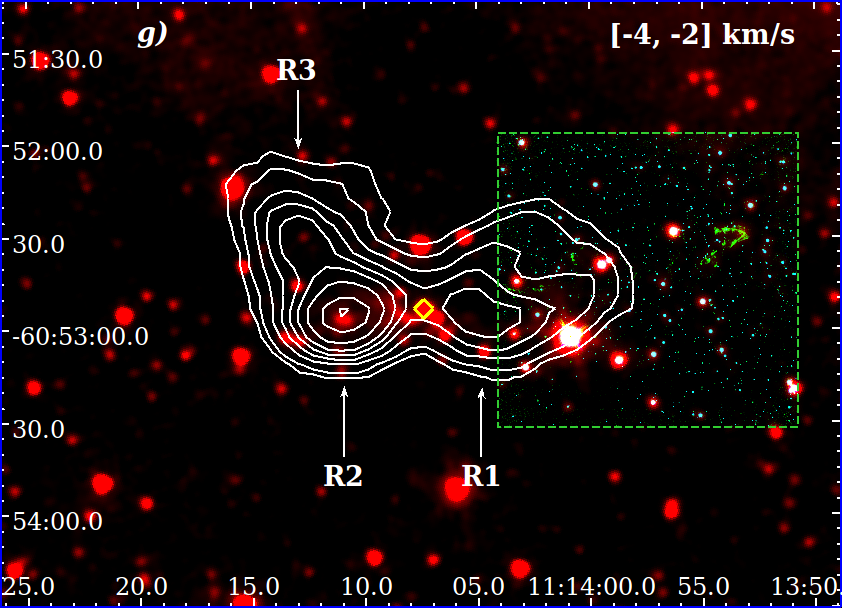}
    \includegraphics[width=0.33\textwidth]{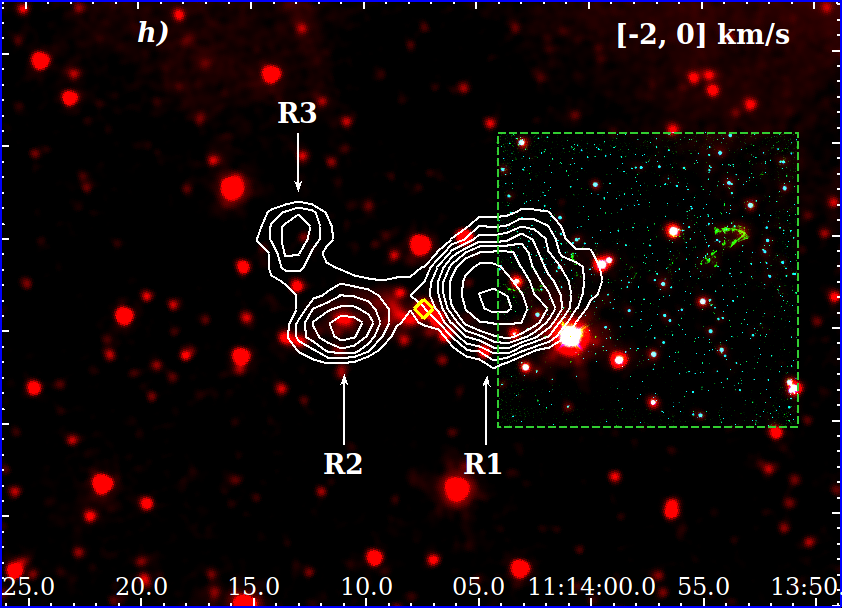}
    \includegraphics[width=0.33\textwidth]{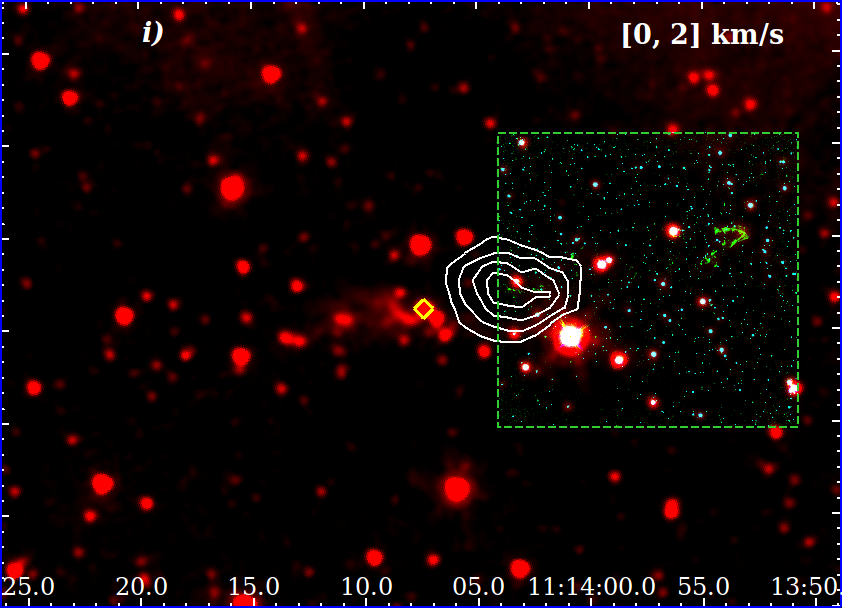}\\
    \caption{$^{12}$CO(3-2) line emissions in the velocity intervals indicated in the upper right corner of each image, superimposed on a combined three-band images: H$_2$ (in green), K (in blue) and 4.5\,\mum\ (in red). All panel have a velocity range of 2~\kms\ except the panel \textit{a)}, which interval is 6~\kms wide. $^{12}$CO(3-2) white contours of panels \textit{a)} to \textit{d)}, \textit{f)} and \textit{g)} correspond to 2, 3, 5, 7, 9, 11, 14, 18, 21 and 21~K~\kms. Panel \textit{e)} shows white contours at 3.5 and 4~K~\kms. Panel \textit{h)} and \textit{i)} display white contours at 1.5, 2, 2.5, 3, 3.5, 4, 5 and 6~K~\kms. The dashed green line box shows the area observed with Gemini. The yellow diamond indicates the location of the suggested exciting source (see Section \ref{sec_driving_source}). In panels \textit{a)}, \textit{b)}, \textit{g)} and \textit{h)} blueshifted (B1, B2 and B3) and redshifted (R1, R2 and R3) structures are marked.}
    \label{fig_12co_contours}
\end{figure*}

\begin{figure*}
    \centering
    \begin{tabular}{ccc}
    \includegraphics[width=0.33\textwidth]{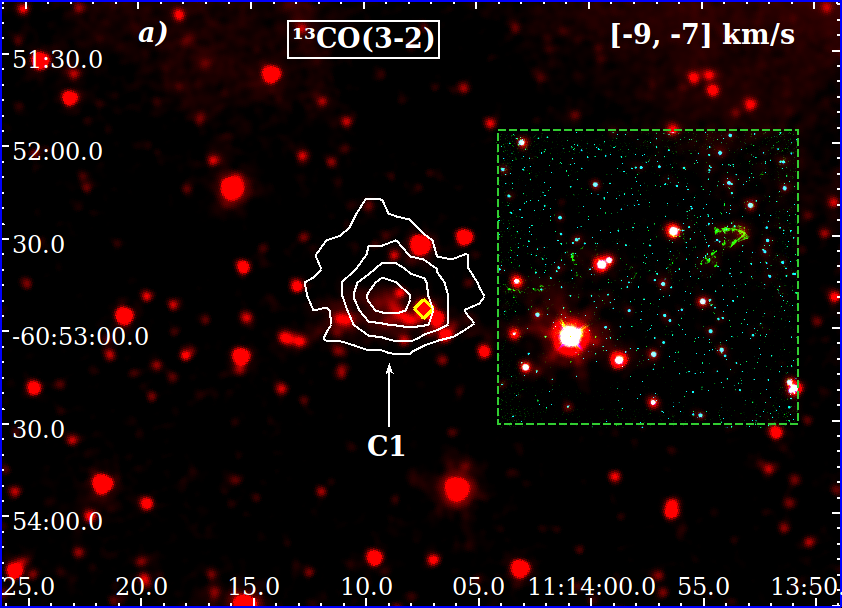}
    \includegraphics[width=0.33\textwidth]{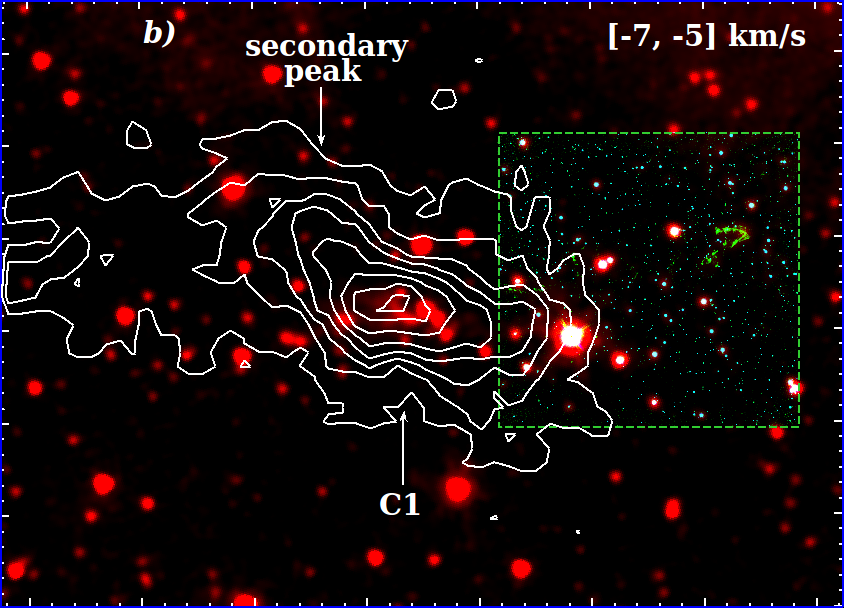}
    \includegraphics[width=0.33\textwidth]{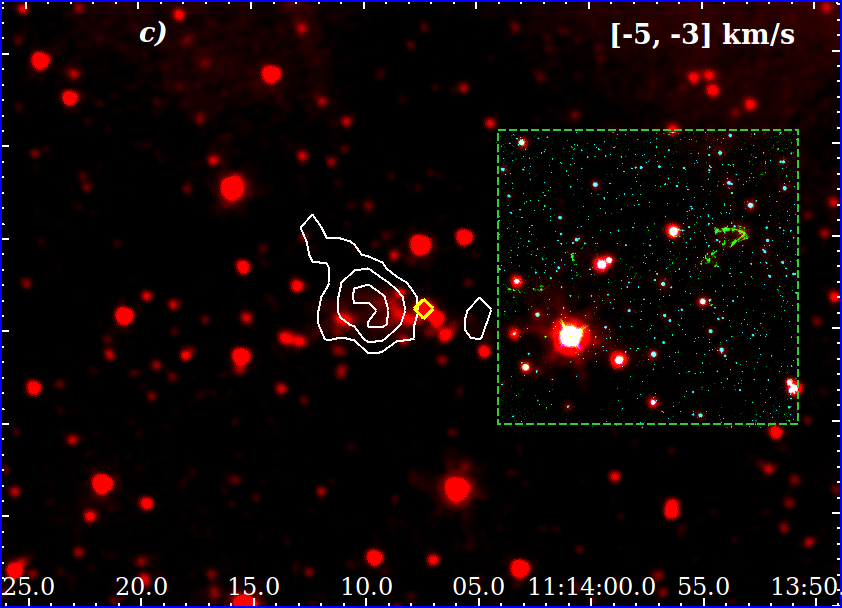}\\
    \includegraphics[width=0.33\textwidth]{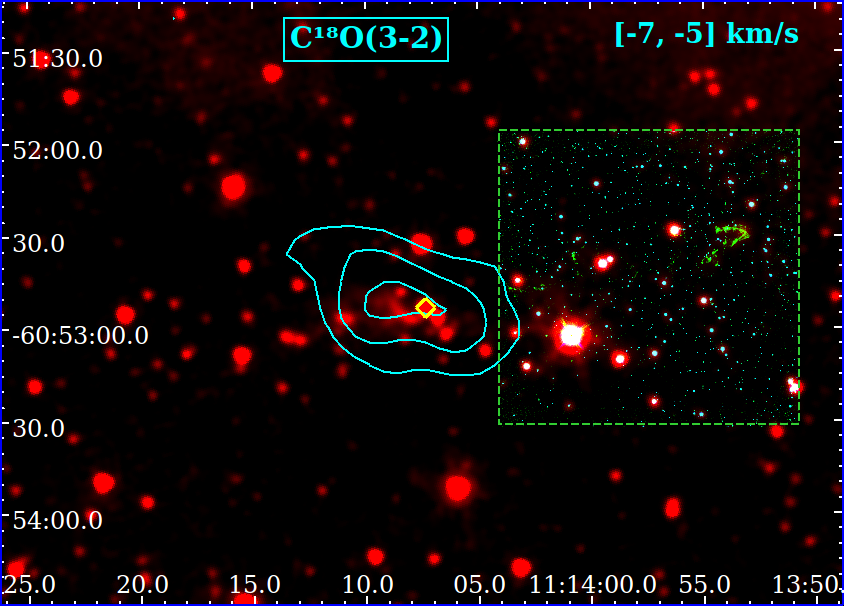} \\  
    \end{tabular}
    \caption{$^{13}$CO(3-2) and C$^{18}$O(3-2) line emissions in the velocity intervals indicated in the upper right corner of each image, superimposed on a combined three-band images: H$_2$ (in green), K (in blue) and 4.5\,\mum\ (in red). All panel have a velocity range of 2~\kms. $^{13}$CO(3-2) white contours (upper panels from \textit{a)} to \textit{c)}) correspond to 2, 3, 4, 5, 6, 7, 8 and 8.8~K~\kms. The lower panel shows C$^{18}$O(3-2) cyan contours at 1, 1.5 and 2~K~\kms. Dense gas coinciding with 
    the positions of the dust clump C1 and a secondary dust peak in Figure~\ref{fig_driving_sources} is indicated. The dashed green line box shows the area observed with Gemini. The yellow diamond indicates the location of the suggested exciting source (see Sec. \ref{sec_driving_source}).}
    \label{fig_13co_c18o_contours}
\end{figure*}

Figures \ref{fig_12co_contours} and \ref{fig_13co_c18o_contours} display the $^{12}$CO, $^{13}$CO and C$^{18}$O integrated emissions at different velocity ranges  overlaid  on a three-color combined image: Gemini K (blue) and H$_2$ (green) and  \Spitzer\ 4.5\,\mum\ (red). The dashed green line box marks the area observed with Gemini. In particular, the $^{12}$CO molecule in Figure~\ref{fig_12co_contours} shows an extended distribution with multiple peak emissions mostly concentrated towards the knots of HH\,137 and HH\,138. It is interesting to note that the position of one of the CO emission peaks, showing an elongated shape (named B2), matches the alignment of all HH\,137 knots, including the H$_2$ terminal bow shock (see panel \textit{d)}). In Section \ref{sec_molecular_outflow}, we analyze such alignment in more detail.

Figure~\ref{fig_13co_c18o_contours} shows the $^{13}$CO and C$^{18}$O integrated emission between --9 and --3~\kms\ in step of 2~\kms.  In this velocity range, we find that the molecular distribution would be associated to a single cloud with an oblate shape orientated from the north--east to the south--west. Since these molecules trace higher densities than the $^{12}$CO, this single cloud will be refereed as a dense clump, coinciding with the dust clump C1 in Figure~\ref{fig_driving_sources}. Nonetheless, when inspecting the $^{13}$CO distribution in more detail (see Figure~\ref{fig_13co_c18o_contours}, panel \textit{b)}), we detect the molecular counterpart of the secondary dust peak shown in Figure~\ref{fig_driving_sources}, detected at $\lambda\,\geq\,160$\,\mum. This second peak is observed between --6.0 and --4.5 \kms. However, we are not able to completely resolve the two $^{13}$CO peaks as different components due to the low resolution of the APEX data.

Fitting an ellipse to the 2.5 K contour ($5\sigma$) of the $^{13}$CO integrated emission (between [--9.5, --3.6]~\kms), we estimate major and minor semi-axes of the molecular clump C1 of $69.6''\times40.3''$ ($\sim 0.51\times0.29$ pc, with a P.A. of 20$^{\circ}$), giving a deconvolved, effective radius $R_{eff} = 0.38$ pc. The center of the fitted ellipse (\hbox{$\alpha =$ 11:14:09.2,} $\delta =$ --60:52:42.700) is located 7$''$ to the north of the maximum emission in $^{13}$CO, and at 20$''$ to the north--east of the suggested exciting source (source 8 in Table~\ref{tab_ysos}). However, these angular separations are practically coincident with the APEX spatial resolution. On the other hand, the C$^{18}$O integrated emission shows a small condensation only between --7 and --5~\kms\ surrounded by the $^{13}$CO emission.

%--------------------FIGURE-5----------------------
%--------------------FIGURE-6-------------------------

%--------------------FIGURE-6-------------------------

% ----------------------------------------------------------------------------------------
\subsection{\texorpdfstring{HCO$^+$(3-2)}{} and HCN(3-2) data}

\begin{figure}
    \centering
    \includegraphics[width=\columnwidth]{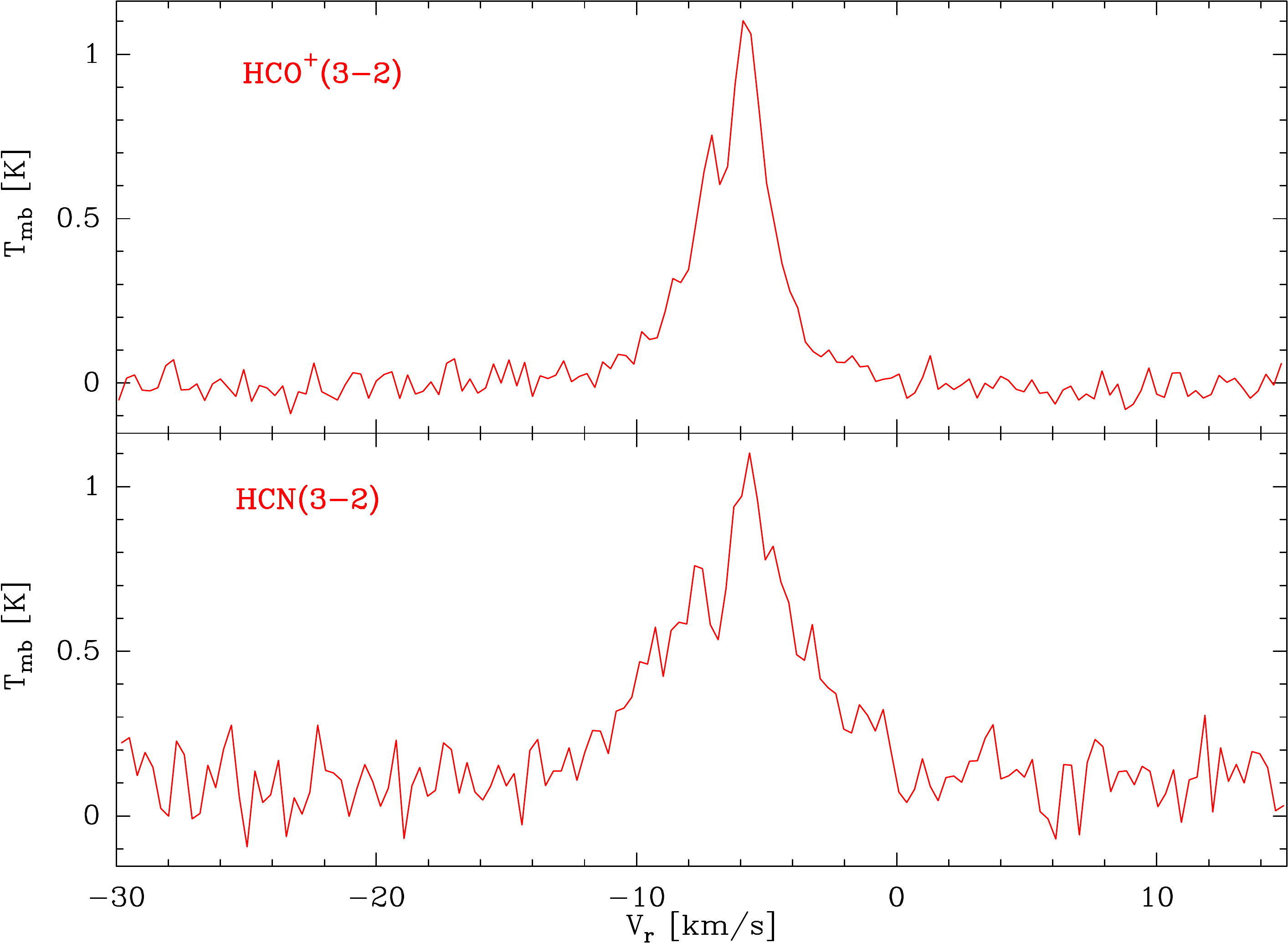}
    \caption{Averaged profiles of HCO$^+$(3-2) and HCN(3-2) within the emitting region of clump C1 in the velocity interval \hbox{[--30, +15]}~\kms~ showing main-beam brightness temperature T$_{mb}$ vs. LSR velocity.}
    \label{fig_hco_hcn_lines}
\end{figure}

Figure~\ref{fig_hco_hcn_lines} shows averaged profiles of HCO$^+$(3-2) (upper panel) and of HCN(3-2) (lower panel) within the emitting region of clump C1. Critical densities for these lines and transitions are: \hbox{$\sim1.5\times10^{6}$\,cm$^{-3}$} and $\sim$ $1.2\times10^{7}$\,cm$^{-3}$, HCO$^+$(3-2) and  \hbox{HCN(3-2)} respectively, for a kinetic temperature T$_k \sim 15$~K \citep{Shirley2015}. Thus, the detection of these lines towards clump C1 indicates ambient densities compatible with these values in the central part of the clump. The HCO$^+$ line has a peak  at \hbox{--5.8~\kms}\ and a weak absorption at --6.8~\kms. The HCN line shows a peak  at --5.7~\kms\ and an absorption, deeper than the HCO$^+$ profile, at $-6.8$~\kms. In addition, the HCN average line profile is wider than the HCO$^+$ profile (6 vs 3~\kms\ at FWHM).

\begin{figure}
    \centering
    \includegraphics[width=\columnwidth]{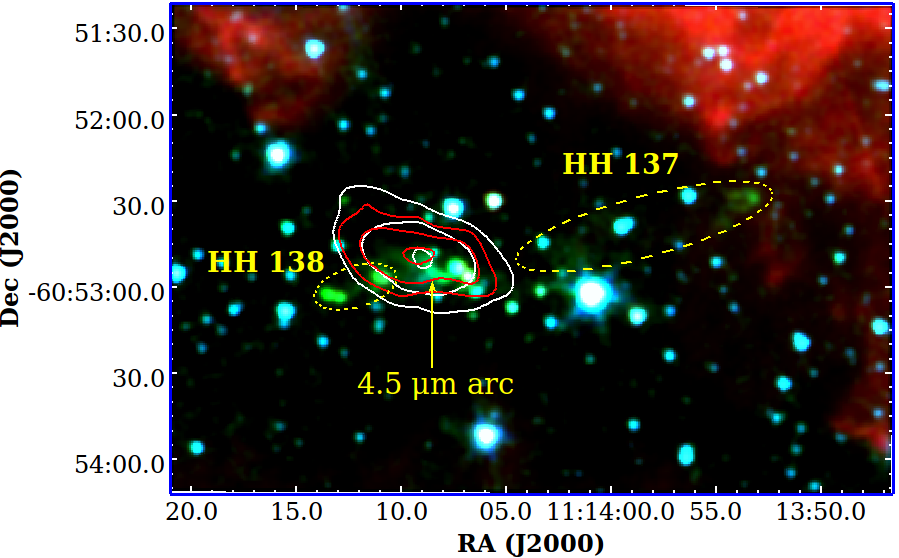}
    \caption{Spatial distribution of HCO$^+$  and  HCN emissions overlaid with the \Spitzer/IRAC images. The white contours are from \hbox{HCO$^+$(3-2)} and correspond to 0.7, 1.2 and 1.7~K. The red contours are from \hbox{HCN(3-2)} and correspond to 0.5, 0.7 and 0.9~K. The combined IRAC images are 3.6\,\mum\ (in blue), 4.5\,\mum\ (in green) and 8.0\,\mum\ (in red). The dashed yellow ellipses show the location of HH\,137 and HH\,138, and the yellow arrow indicates the 4.5\,\mum\ arc.}
    \label{fig_hco_hcn}
\end{figure}

Figure~\ref{fig_hco_hcn} shows the spatial distribution of the HCO$^+$ and HCN emissions overlaid with the IRAC/\Spitzer\ 3.6, 4.5 and 8.0\,\mum\ images. In HCO$^+$, the elongated emitting region is $56\arcsec~\times~30\arcsec$ (0.41\,pc~$\times$~0.22\,pc), and it is detected in the interval \hbox{[--11.8, --2.9]~\kms}. In the case of HCN, the emission is present in a region of $42\arcsec~\times~21\arcsec$ (0.31\,pc~$\times$~0.15\,pc) in the velocity range \hbox{[--12.3, 0.4]~\kms}. These sizes are estimated from contours at 5$\sigma$ in the integrated emission. In both lines, the emission coincides with the brightest part of C1 and the cold dust emission (see Figure \ref{fig_13co_c18o_contours}). In particular, the arc at 4.5\,\mum\ is projected onto the brightest regions, while knots A and B belonging to HH\,138 are within the molecular region as well as the proposed driving source (see upper panel of Figure \ref{fig_spitzer_egos}). 

%---------------------FIGURE-9-----------------------

%--------------------FIGURE-10--------------------
% ----------------------------------------------------------------------------------------
\subsection{Physical parameters of the molecular core} \label{sec_parameters_gas}

We estimated the H$_2$ mass of clump C1 assuming LTE condition through both $^{13}$CO and C$^{18}$O emissions. We assumed that the excitation temperature is the same for both molecules with a value of T$_{\text{exc}}=15$~K.
To fix this value, we calculated the excitation temperature from the $^{12}$CO maximum giving T$_{\text{exc}}=13.4$~K, close to the dust temperature T$_{\text{dust}}=14.8$~K (see Section \ref{sec_infrared}). Nevertheless, the derivation of T$_{\text{exc}}$ from the  $^{12}$CO maximum was not completely reliable due to the self-absorption of this line (see Figure~\ref {fig_perfil_co}, upper panel). However, this T$_{\text{exc}}$ should not be too
high since the characteristic excitation temperature for IRDCs
is $\sim$ 10 K \citep[see][]{Du2008}. Thus, we assumed for this region an excitation temperature close to the dust temperature of 15~K.

 We followed the equations of \cite{Bourke1997} to derive the optical depths ($\tau_{13}$, $\tau_{18}$), column densities and the molecular mass. The optical depths of $^{13}$CO and C$^{18}$O were estimated using the following equation:

\begin{equation}
\tau=-\text{ln}\left[1-\frac{T_{\text{peak}}}{T^*}
\left[\left(e^{\frac{T^*} {T_{\text{exc}}}}-1
\right)^{-1}-\left(e^{\frac{T^*} {T_{\text{bg}}}}-1\right)^{-1}\right]^{-1}\right], 
\label{tau13co}
\end{equation}

\noindent
where,  $T^*$ = $h \nu / k$, with $\nu$  the rest frequency of the $^{13}$CO(3-2) and C$^{18}$O(3$-$2) lines and $T_{\text{bg}}$ the background temperature of 2.7 K. To obtain the main-beam brightness temperature peaks for both molecular lines ($T_{\text{peak}}^{13}$, $T_{\text{peak}}^{18}$), we performed a Gaussian fit to the $^{13}$CO and C$^{18}$O averaged spectra (Figure~\ref{fig_perfil_co}) within the area of the molecular clump (132\arcsec $\times$ 81\arcsec, with a P.A. of \,$20\degr$). The $^{13}$CO and C$^{18}$O molecular lines have peak temperatures of $\sim$ 2.0 and $\sim$ 0.55 K, respectively. However, since the  $^{13}$CO profile is weakly affected by a self-absorption (see Figure~\ref{fig_perfil_co}, middle panel), we adopted a $T_{\text{peak}}^{13} \approx 2.5$~K.

With these peak temperatures, we estimated that $ \tau_{13}$ and  $\tau_{18}$ are  0.33 and 0.07, respectively, showing that both lines are optically thin. Then, the $^{13}$CO column density was estimated using: 

\begin{equation}
N({\rm^{13}CO})=8.07\times 10^{13}  e^{\frac{15.86}{T_{\text{exc}}}} \left[\frac{T_{\text{exc}}+0.88}{1-e^{-\frac{15.86}{T_{\text{exc}}}}} \right] \int \tau^{13}\   d\upsilon \ \  \textrm{ (cm$^{-2}$)}
\label{n13co}
.\end{equation}
Since the $^{13}$CO line is optically thin, the integral of Eq.~\ref{n13co} can be replaced by
\begin{equation}
\int{\tau^{13} d\upsilon \approx\ \frac{1}{J{(T_{\text{ex}})}-J{(T_{\text{bg}}})} \int{T_{\text{mb}}}}\ \ d\upsilon,
\label{integral}
\end{equation}
with
\begin{equation}
    J(T)=\frac{T^*}{e^{\frac{T^*}{T_{\text{ex}}}}-1} .
\end{equation}

\indent 
The total molecular hydrogen mass is calculated using: 

\begin{equation}\label{eq:masa1}
\text{M(H$_2$)}\ =\ \mu\ m_H\  A \ N(\text{H}_{2})\ d^2 \quad \quad \quad \textrm{($M_\odot$)}
,\end{equation}

\noindent
where $\mu$ is the mean molecular weight, equal to 2.76 for a relative helium abundance of 25\% by mass \citep{Yamaguchi1999}, m$_{\text{H}}$ is the hydrogen atom mass (1.67$\times 10^{-24}$~g), $A$ is the area subtended by the $^{13}$CO clump, N(H$_2$) the total molecular column density, and $d$ is the adopted distance (1.5\,$\pm$\,0.5~kpc). 
To obtain the N(H$_2$) and the total molecular hydrogen mass, an abundance of [${\text{H}_2}$/$^{13}$CO]~=~$1\times10^6$ \citep{Saldanio2019} was adopted. Our result for the clump C1 turns out to be N(H$_2$)~$=4.3\times10^{21}$~cm$^{-3}$, and M(H$_2$)~$=$~36~\msun, within an area with an effective radius of $R_{eff} =$~0.38 pc. The uncertainty in the column density is about 20\%. Such error originates mainly in the inaccuracy of the distance. In addition, the unknown geometry of the source and the uncertainty in abundances also contribute to the estimated error. With regard to the mass, if we considered T$_{\text{exc}}=15 \pm 5$~K, the mass would turn out to be M(H$_2$)~$=~36 \pm 29$~\msun.

%=========================================================================================
\section{The molecular outflows}  \label{sec_molecular_outflow}

As mentioned in Section \ref{sec_co_data}, the $^{12}$CO(3-2) profile shows noticeable wings at lower and higher velocities with respect to the systematic velocity.  We applied the criterion of \cite{Hatchell2007}\footnote{These authors provide an objective criterion to identify sources with outflows by measuring the $^{12}$CO(3$-$2) line--wings with more than $3\sigma$ at 3 \kms\ from the C$^{18}$O(3-2) line center.} to identify both redshifted and blueshifted components (red lobes and blue lobes hereafter) between \hbox{[--20, --9.3]~\kms}\ and \hbox{[--3.3, +5]~\kms}, respectively. These velocity ranges are indicated with dashed red and blue line boxes in Figure~\ref{fig_perfil_co}.

\begin{figure*}
     \centering
     \includegraphics[width=12cm]{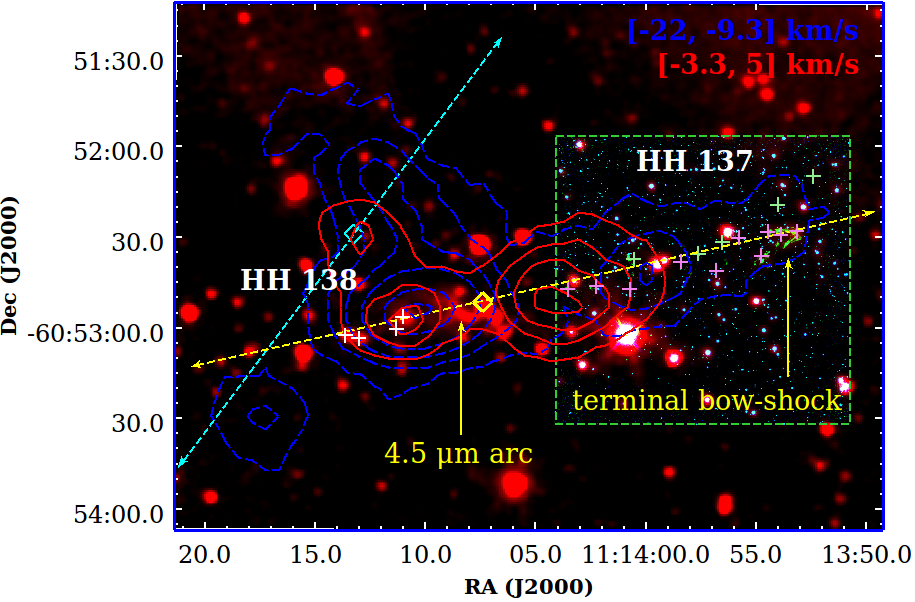}\\
     \includegraphics[width=7.4cm]{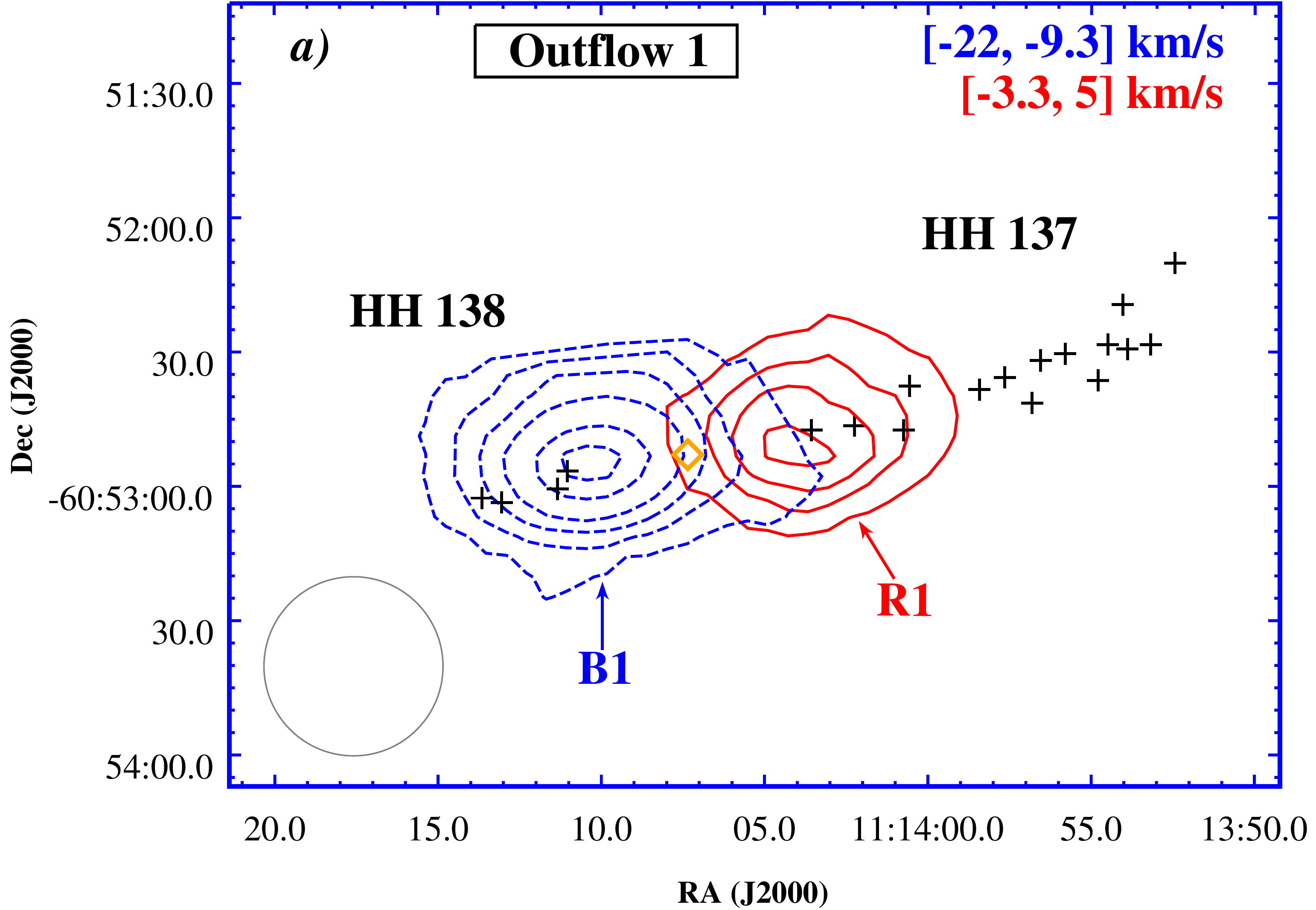}
     \includegraphics[width=7.4cm]{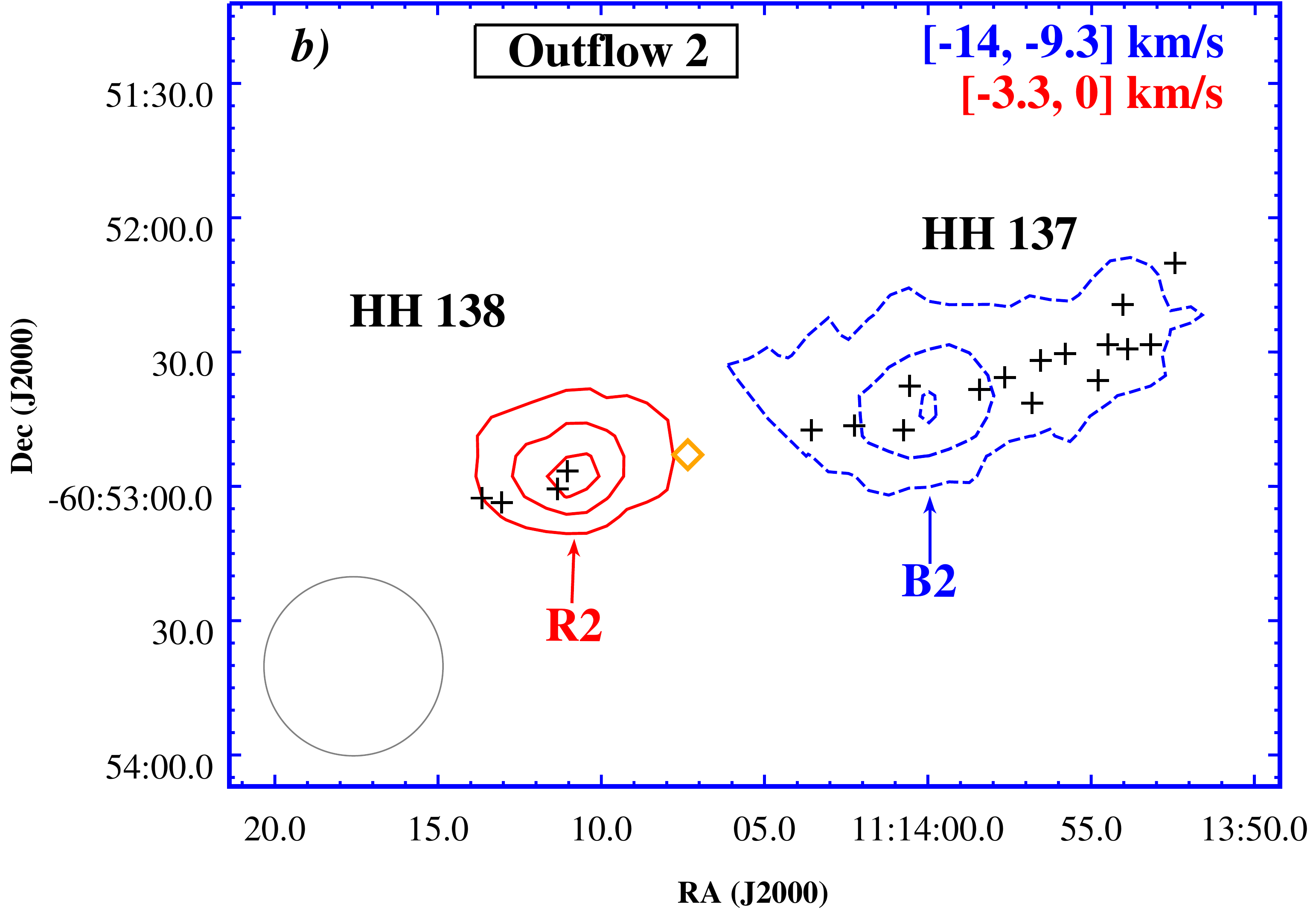}
     \includegraphics[width=7.4cm]{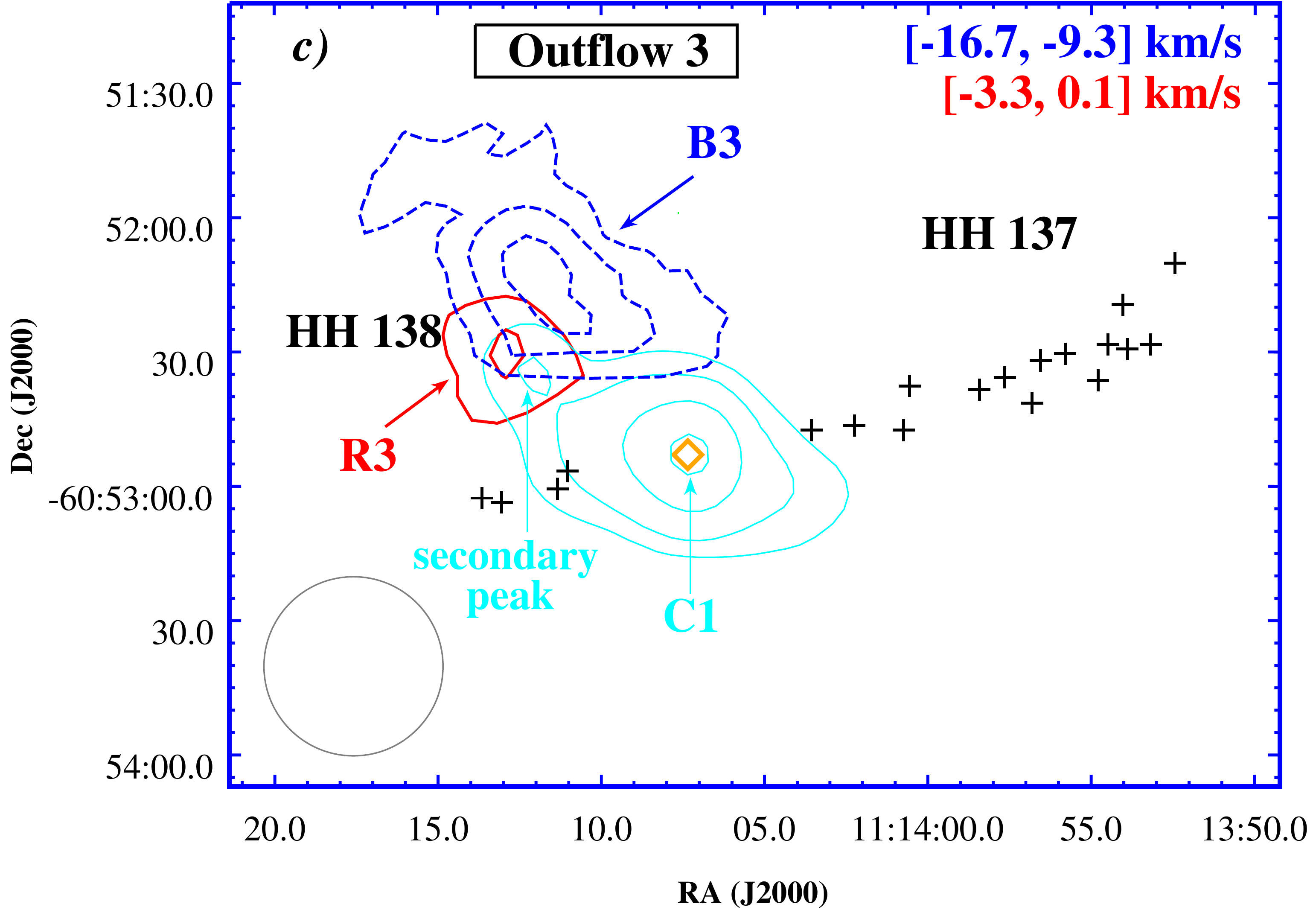}
     \caption{\textit{Upper panel:} Blueshifted and redshifted wings of the $^{12}$CO molecular line emission in the velocity ranges of \hbox{[--22, --9.3]}~\kms\ and \hbox{[--3.3, 5]}~\kms, respectively, superimposed on a combined three band images: K (blue), H$_2$ (green), and 4.5\,\mum\ (red). The dashed blue lines correspond to 2, 6 12, 20, 34 and 44~K~\kms\ and the continuous red line to 4, 8, 11, and 14~K~\kms. The magenta and light green crosses indicate the positions of optical knots of HH~137 \citep{Ogura1993} and new H$_2$ knots from Section \ref{sec_gemini}. The white crosses mark the optical knots of HH~138 from \citet{Ogura1993}. The dashed green line box underlines the field observed with Gemini. The terminal H$_2$ bow shock as well as the 4.5\,\mum\ arc--like structure are indicated with yellow arrows. The double arrow dashed yellow line marks the cutting line used to make the Position--Velocity diagram in Figure~\ref{fig_PV_outflow}, upper panel, indicating the outflow direction from coordinates \hbox{\radec$~=~$(11:13:49.605,--60:52:21.69)} to \hbox{(11:14:20.693, --60:53:13.05)}. The yellow diamond marks the location of the \Spitzer\ and \WISE\ exciting sources discussed in Section \ref{sec_driving_source}, chosen as the center of offset positions shown in Figure~\ref{fig_PV_outflow} (upper panel). The double arrow dashed cyan line indicates the cutting line used to make the Position--Velocity diagram in Figure~\ref{fig_PV_outflow}  (lower panel), showing the outflow direction from coordinates \hbox{\radec$~=~$(11:14:21.222, --60:53:46.48)} to \hbox{(11:14:06.512, --60:51:23.76)}. The cyan diamond marks the position of the offset center in Figure~\ref{fig_PV_outflow}, lower panel, located at \hbox{\radec$~=~$(11:14:12.054, --60:52:35.50)}. The center of this offset position lies 10.8~\arcsec\ north--west of the secondary dust peak in Figure~\ref{fig_driving_sources}. 
     \textit{Middle and bottom panels:} Blueshifted and redshifted lobes associated to the outflow 1 (middle left panel), outflow 2 (middle right panel) and outflow 3 (bottom panel) in the velocity ranges indicated in the upper right corner of each panel. The orange diamond marks the location of the suggested exciting stars (see Section \ref{sec_driving_source}). In the bottom panel \textit{c)}, the cyan contours are the 160\,\mum\ dust emissions associated with clump C1 and a secondary dust peak (see Figure  \ref{fig_driving_sources}).
     The HPBW is shown in the lower left corner of each panel.}
     \label{fig_outflow_wings}
 \end{figure*}

\begin{figure*}
    \centering
    \includegraphics[width=\textwidth]{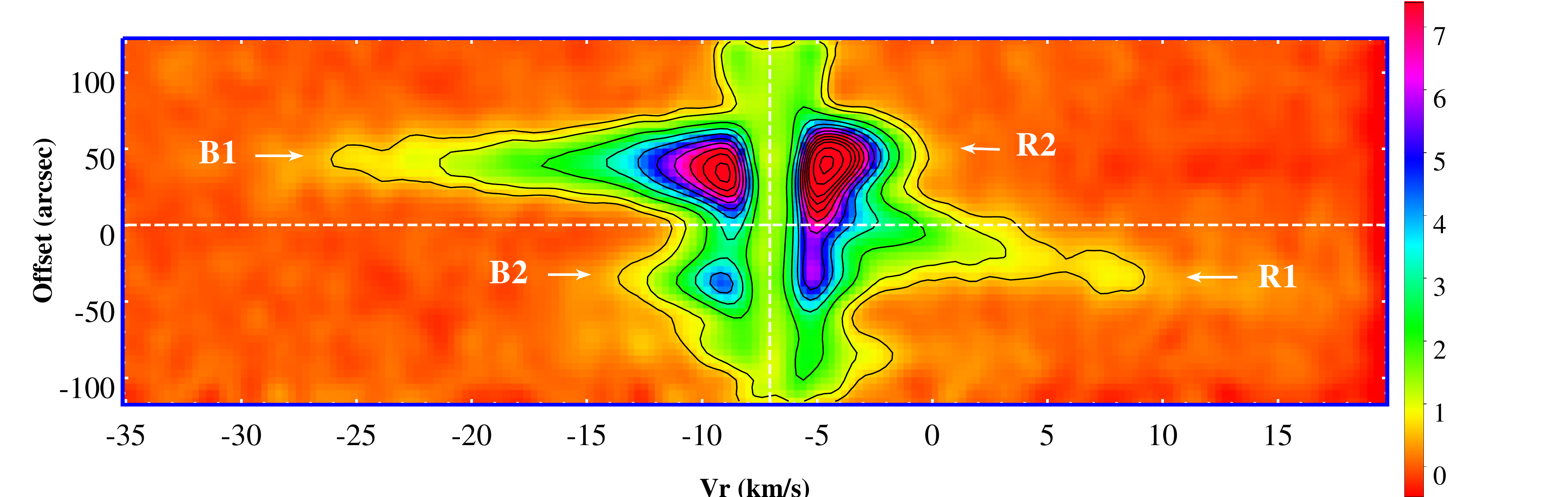}
    \includegraphics[width=\textwidth]{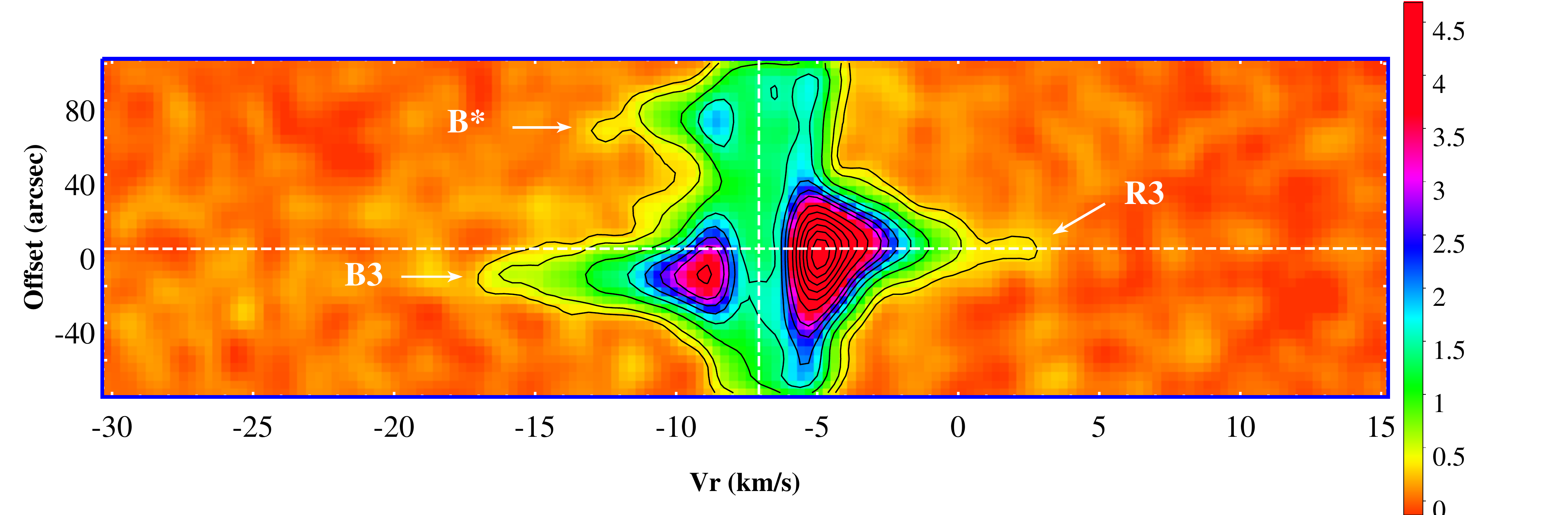}
    \caption{Position--Velocity (PV) diagram of the $^{12}$CO(3-2) emission along two transverse cuts shown in Figure~\ref{fig_outflow_wings}. 
    \textit{Upper panel:} Cut along the double arrow dashed yellow line in Figure~\ref{fig_outflow_wings}. Offset coordinates are referred to the suggested exiting sources positions (see Section \ref{sec_driving_source}), located at \radec~=~(11:14:07.397, --60:52:51.39). The black contours correspond to 0.6, 1, 2, 3, 4, 5, 6, 7, 8, 9, 10, 12 and 14~K.
    \textit{Lower panel:} Cut along the double arrow dashed cyan line in Figure~\ref{fig_outflow_wings}. Offset coordinates are referred to a position (\radec~=~(11:14:13.240, --60:52:35.29.03)) on the cutting line, indicated with a cyan diamond, 10.8~\arcsec\ to the north--west of the position of the secondary peak. The black contours correspond to 0.3, 0.5, 1, 1.5, 2, 2.5, 3, 4, 5, 6, 7, 8 and 9~K. The blueshifted emission identified as B*, corresponds to the blue contours at the south--east corner in Figure~\ref{fig_outflow_wings}. The color scale is indicated at the right. The horizontal dashed white line marks the zero offset position. The vertical dashed white line corresponds to the velocity of the central depression detected in the $^{12}$CO molecular line (see Figure~\ref{fig_perfil_co}).}
    \label{fig_PV_outflow}
\end{figure*}

Figure~\ref{fig_outflow_wings}, upper panel, shows the $^{12}$CO line--wings integrated emission overlaid on a combined three-color image, K(blue), H$_2$(green) and 4.5\,\mum\ (red), showing that four lobes (two red and two blue) are aligned with the knots of HH\,137 and HH\,138. Performing a cut along the knots direction, as is indicated in the upper panel of Figure~\ref{fig_outflow_wings} with a double arrow dashed yellow line, two blue lobes (named B1 and B2) and two red lobes (R1 and R2) are identified in the position-velocity (PV) diagram of Figure~\ref{fig_PV_outflow}. The cut goes from \radec~=~(11:13:49.605, --60:52:21.69) to (11:14:20.693, --60:53:13.05), with offset coordinates centered on the proposed exciting source position indicated with a horizontal dashed white line in this figure (see Section  \ref{sec_driving_source}). Panels a) and b) of Figure~\ref{fig_outflow_wings} show $^{12}$CO contours emissions for these two outflows (outflows 1 and 2).

The B1 lobe, as is shown in Figures \ref{fig_outflow_wings} (middle panel \textit{a)}) and \ref{fig_PV_outflow} (upper panel), is observed from --22 to --9.3~\kms, while the B2 lobe  appears from --14~\kms. The elongated shape of B2, which coincides with all H$_2$ knots of HH\,137, ends at the terminal bow shock to the north--west. Meanwhile, B1 is located towards \cite{Ogura1993}'s knots of HH\,138, as well as superimposed to the R2 red lobe. This later lobe is less extended in velocity (from --3.3 to 0~\kms). On the other hand, the R1 red lobe is more extended in velocity than R2 (from --3.3 to +5~\kms) and overlays with the knots of HH\,137 that lie closer the proposed exciting source (see Section \ref{sec_driving_source}).  Asymmetries in the PV diagram similar to those in Figure~\ref{fig_PV_outflow}, upper panel, have been previously observed in other bipolar outflows analyzed using data of similar spatial resolution to the APEX data \citep[e.g.,][]{Bourke1997,Sanhueza2010}. In view of this, we associate the B1 blue lobe with the R1 red lobe, and similarly the B2 lobe to the R2 lobe. These associations are indicated in panels a) and b) of Figure~\ref{fig_outflow_wings}. 

Our description of the blueshifted and redshifted lobes is consistent with a scenario in which the driving sources, nested in the dense clump C1, develop two molecular outflows projected along the H$_2$ knots. As was shown in Figure~\ref{fig_driving_sources}, at least two \Spitzer\ Class I sources (YSO 1 and 2 in Table~\ref{tab_ysos}) lying towards the dust clump were found.  As discussed in Section \ref{sec_driving_source}, YSO 1 and 2 have a projected separation of $\sim 7500$~AU and might form a wide binary system, being each star the exciting source of each outflow.

Multiple  bipolar molecular outflows are commonly associated with multiple protostellar systems sharing the same  parental envelope \citep[e.g.,][]{Wu2009,Lee2016}. In addition, the orientations of the outflows axes may provide some hint on the directions of multiple protostellar rotation or spin axes.  Different proposed scenarios for the formation of aligned binary or multiple systems have been enumerated by  \cite{Lee2016}. These scenarios foresee the following: a) formation in a massive disk or ring in which two or more  large co-rotating structures appear, b) binary components born by fragmentation of a core which originates two or more central objects with aligned angular momentum  vectors, c) tidal effects that bring together two or more protostellar objects during subsequent evolutionary phases.

In the case of the outflows in Figures \ref{fig_PV_outflow}, panels a) and b), the inclination angles with respect to the plane of the sky should be small since both the red and blue lobes are well resolved and the centers of each lobe pair are separated by $\sim~47''$ (outflow 1) and 81\arcsec\ (outflow 2), higher than APEX HPBW of $\sim$ 20\arcsec.  These outflows show similar P.A. (with an angular difference of about $\sim 8^{\circ}$). Nonetheless, the projected angular moments of both outflows are probably in opposite senses. Hence, assuming that these outflows are associated with a binary system, the exciting stars would have almost aligned but opposite direction rotation axes, which may favor the later tidal scenario for the binary pair formation.

\begin{figure*}
    \centering
    \includegraphics[width=0.85\textwidth]{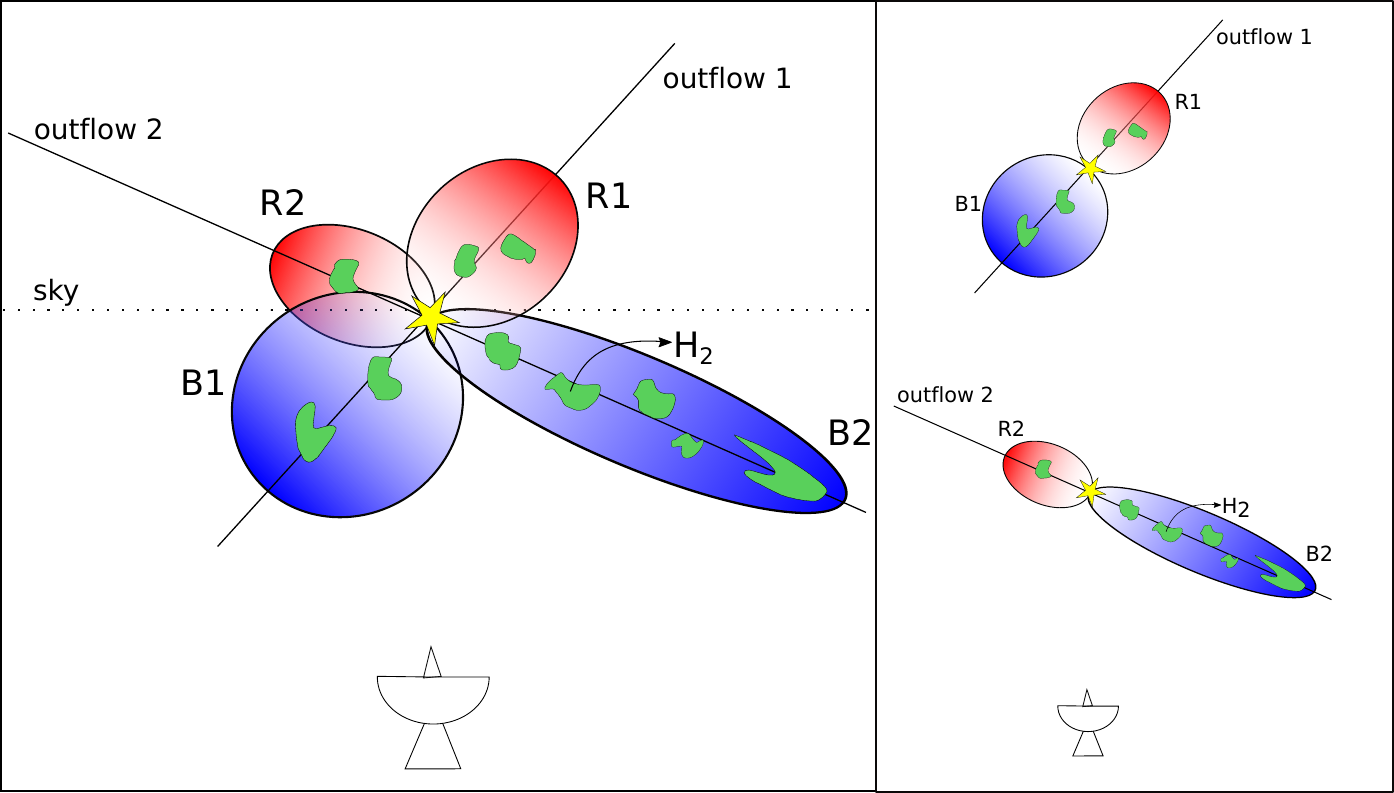}
    \caption{Schematic representation of outflows 1 and 2 with respect to the plane of the sky. In the left panel, the outflows are at the same distance and the driving sources form a binary star. In the right panel,x the outflows are at different distances and no physical association between the exciting stars exists.}
    \label{fig_scheme}
\end{figure*}

Figure~\ref{fig_scheme} shows a schematic representation of the positions of outflows 1 and 2 with respect to the plane of the sky. In the left panel, the outflows are at the same distance and the driving sources form a binary star whereas  in the right panel, the outflows are at different distances  and no physical association between the exciting sources exits.  Thus, the outflows are seen projected on the plane of the sky. Observations at higher spatial resolution, like those that can be performed by the Atacama Large Millimeter/submillimeter Array (ALMA) telescope, are needed to shed light on the outflows alignment and the binary nature of the exciting source. 

On the other hand, a third likely weak outflow may be identified to the north-east of the four-lobe system (outflows 1 and 2). The blue and red lobes, named as B3 and R3 in the panel c) of Figure~\ref{fig_outflow_wings}, are observed between \hbox{[--16.7, --9.3]}~\kms\ and \hbox{[--3.3, 0.0]~\kms}, respectively. These lobes are projected towards the weaker secondary peak, observed at wavelength $\geq 160$~\mum\ (see Figure~\ref{fig_driving_sources}). Figure~\ref{fig_PV_outflow}, lower panel, shows the PV diagram of outflow 3. The cut along this outflow is indicated with double arrow dashed cyan line in Figure~\ref{fig_outflow_wings} and goes from \hbox{\radec~$=$~(11:14:21.222, --60:53:46.48)} to \hbox{(11:14:06.512, --60:51:23.76)}. Offset positions are centered 10.8\arcsec\ to the north--west of the secondary peak and indicated with a horizontal dashed white line in the lower panel of Figure~\ref{fig_PV_outflow}. If the third outflow was associated with the secondary peak, this condensed structure would be a very young object surrounded by a cold and thick envelope completely shielding the exciting proto-star. The lack of \WISE\ and \Spitzer\ young candidate sources near the secondary peak (see Figure~\ref{fig_driving_sources}) might provide support to this suggestion. Figure~\ref{fig_PV_outflow}, lower panel, also shows blueshifted emission identified as B*, corresponding to the blue contours at the south--east corner in Figure~\ref{fig_outflow_wings}.

Several physical parameters of the outflows can be estimated following, for example, \cite{Beuther2002,Yang2018} and \cite{deVilliers2014}. They provide information about energy and mass that help to characterize the outflow. The parameters obtained from the molecular observations are: the length $r$ of the jet, the integrated emission in $^{12}$CO(3-2) along the velocity range, the mass of the outflow $M_{out}$, the momentum $p$, the mechanical energy $E$, the time scale $t$, the mass entertainment rate $\dot{M}_{out}$, the mechanical force $F_m$, and the mechanical luminosity $L_m$:

\begin{align}
    p =& \ M_{out} \times \ \text{v}_{max-out}\ ,         \\
    E =& ~ \frac{1}{2} \ M_{out} \times \ \text{v}_{max-out}^{2}\ ,\label{mout}\\ 
    t =& ~ \frac{r}{\text{v}_{max-out}}\ ,               \\
    \dot{M}_{out} =& ~ \frac{M_{out}}{t}\ ,     \\
    F_{m} =&  ~\frac{p}{t}\ ,                   \\
    L_{m} =& ~\frac{E}{t}\ ,
\end{align}

\noindent
where v$_{max-out}$ is the maximum velocity in the blue and red lobes, respectively. Bearing in mind that the whole outflows are detected in the $^{12}$CO(3-2) line only (with exception of a small core, which is seen in the $^{13}$CO(3-2) and C$^{18}$O lines), we estimated the H$_2$ mass using the relation between the H$_2$ column density and the $^{12}$CO integrated emission. Taking into account the solid angle of the outflows (which give the linear sizes, r, in Table \ref{tab_parameters}), the corresponding H$_2$ masses amount to $M_{out} =$ 17, 5 and 4~\msun, outflows 1, 2 and 3. 

Table~\ref{tab_parameters} lists these parameters for the three outflows in Figures \ref{fig_outflow_wings} and \ref{fig_PV_outflow}. The values given in this table are not corrected by the unknown inclination angle. However, as mentioned before in this section, in the cases of outflows 1 and 2, red and blue lobes are well resolved and the corresponding centers have separations larger than APEX data HPBW, which indicates that the inclination angles should be small. The length of the outflow 2 from the terminal bow shock to the proposed driving source is $0.9\pm0.3$~pc. However, this parameter depends on the inclination of the outflow with respect to the plane of the sky.

\cite{Downes2007} suggest that no inclination correction is needed for the momentum, since, by coincidence, underestimates along the jet axis are canceled by overestimates in the transverse component. On the other hand, the energy is overestimated by the inclination angle.  Correction factors propose by these authors are applied to jet driven outflows emanating from Class 0 sources than have not yet broken out the parent cloud, which is not the case of outflows 1 and 2.

The parameters of the outflows identified in the HH\,137 and HH\,138 region are, in general, quite similar to those reported by \cite{Beuther2002} in high-mass star-forming regions and by \cite{Yang2018} towards ATLASGAL clumps. The first authors used the \hbox{$^{12}$CO $J=$ 2--1} transition to identified outflows whereas in the paper of \cite{Yang2018} the \hbox{$^{13}$CO(3-2)} and C$^{18}$O(3-2) lines were employed. Thus, the parameters the outflows associated with HH\,137 and HH\,138 are typical of high-mass stars.

%---------------------FIGURE-7-----------------------
%------------------------FIGURE-8---------------
%-----------------------------------------------
%------------------------------TABLE-4-------------------  
\begin{table*}
\centering
\caption{Parameters of the outflows.}
\label{tab_parameters}
\begin{tabular}{lcccccc}
\hline
  &  \multicolumn{2}{c}{{Outflow 1}} & \multicolumn{2}{c}{{Outflow 2}} & \multicolumn{2}{c}{{Outflow 3}}\\
  Parameters  &   B1  &   R1  &   B2  &   R2  &   B3   & R3 \\
\hline
 Velocity range [km s$^{-1}$]	& $[-22,\,-9.3]$    & $[-3.3,\,5]$      & $[-14,\,-9.3]$    & $[-3.3,\,0]$    & $[-16.7,\,-9.3]$     & $[-3.3,\,0.1]$  \\  
 $\int I_{^{12}CO} dv$ [K km s$^{-1}$]  & 2272~$\pm$~48 & 924~$\pm$~30  & 654~$\pm$~26  & 421~$\pm$~21 & 658~$\pm$~26  & 174~$\pm$~13  \\
 v$_{max}$ [\kms]               & 15.7~$\pm$~0.3  &   11.3~$\pm$~0.3 & 7.7~$\pm$~0.3   & 5.8~$\pm$~0.3 &  10.4~$\pm$~0.3     &   6.4~$\pm$~0.3 \\
 r [pc]          				& 0.4~$\pm$~0.2   &  0.4~$\pm$~0.2   & 0.9~$\pm$~0.3   & 0.3~$\pm$~0.2 & 0.7~$\pm$~0.2      &   0.4~$\pm$~0.2 \\
 $M_{out}$ [\msun] 	    	    & 12~$\pm$~8      &   5~$\pm$~3     & 3~$\pm$~2       & 2~$\pm$~1     & 3~$\pm$~2          &   0.9~$\pm$~0.6 \\
 $p$ [\msun\,\kms]              & 182~$\pm$~122   &  53~$\pm$~36    & 26~$\pm$~17     & 12~$\pm$~8    & 35~$\pm$~23        &   6~$\pm$~4 \\
 $E$ [10$^{45}$ erg]		    & 28~$\pm$~19     &  6~$\pm$~4      & 2~$\pm$~1       & 0.7~$\pm$~0.5 & 4~$\pm$~2          &  0.4~$\pm$~0.2  \\
 $t$ [10$^{4}$ yr] 				& 3~$\pm$~1       &  4~$\pm$~2       & 12~$\pm$~4      & 6~$\pm$~3    & 7~$\pm$~2          &  7~$\pm$~3 \\
 $\dot{M}_{out}$ [10$^{-5}$ \msun\,yr$^{-1}$]       &  41~$\pm$~34     & 12~$\pm$~10      & 3~$\pm$~2    & 3~$\pm$~3           & 5~$\pm$~4       & 1~$\pm$~1 \\
 $F_{m}$ [10$^{-4}$ \msun\,\kms\,yr$^{-1}$]         &  65~$\pm$~53     & 13~$\pm$~11     & 2~$\pm$~2    & 2~$\pm$~2           &  5~$\pm$~4       & 0.8~$\pm$~0.6 \\
 $L_{m}$ [L$_{\odot}$] 		    & 8~$\pm$~7      & 1~$\pm$~1    & 0.1~$\pm$~0.1 & 0.09~$\pm$~0.08 & 0.4~$\pm$~0.3   &  0.04~$\pm$~0.03 \\ 
\hline
\end{tabular}
\end{table*}
%-----------------------------------------

%-----------------------------------------------
%-----------------------------------------------

%=========================================================================================

\section{Summary} \label{sec_summary}

In this contribution, we present a multiwavelength study of the Herbig-Haro objects 137 and 138 identified by \cite{Ogura1993} in the Carina region. High resolution Gemini H$_2$ images allowed us to detect 2.12\,\mum\ shock emissions linked to most of the optical knots composing the chain of emissions in HH\,137, as well as five new H$_2$ knots. These new H$_2$ emissions are designated as MHO~1629 in  the on-line Catalogue of Molecular Hydrogen Emission-Line Objects \citep[MHOs,][]{Davis2010}. We inspected \Spitzer\ 4.5\,\mum\ images finding counterparts for almost all H$_2$ knots reported in this work. We also identified new likely shock excited regions associated with the four optical knots delineating HH\,138. 
Moreover, a bright 4.5\,\mum\ 0.09~pc-long arc-shaped structure, roughly located mid-way between HH\,137 and HH\,138 was found. This is likely associated with two \Spitzer\ point-like sources located very close to the arc, which coincides with a peak of emission in 70 and 160\,\mum\ and with the \hbox{WISE~J111406.96--605255.9} source.  The arc-shaped structure, as well as the two point-like sources, are projected on the axis of HH\,137 and HH\,138. We suggest that these stars are the exciting sources of these HH objects.

$^{12}$CO(3-2), $^{13}$CO(3-2) and C$^{18}$O(3-2) molecular line data obtained with the APEX telescope allow us to identify a dense core (C1) for which we derive a LTE mass of 36~\msun. This core, also detected in the high density HCO$^+$(3-2) and HCN(3-2) molecular lines, coincides with the position of the \WISE\ candidate exciting source.

The $^{12}$CO(3-2) emission distribution over the observed region reveals molecular material associated with at least three outflows. Outflows 1 and 2 extend along the chain of optical knots forming HH~138 and HH~137 discovered by \citet{Ogura1993} and Outflow 3 lies to the north--east.
In addition, outflows 1 and 2 partially overlap on the plane of the sky. The appearance of these outflows on the plane of the sky suggests a scenario in which the driving sources, nested in the dense clump C1, develop two molecular outflows projected along the optical and H$_2$ knots. \Spitzer\ Class I sources (YSO 1 and 2) lie towards the dust clump (C1), have a projected separation of $\sim 7500$~AU and might form a wide binary system, being each star the exciting source of each outflow. Outflow 3 is projected towards a secondary weaker dust clump observed at wavelength $\geq 160$~\mum.

We estimate H$_2$ masses $M_{out} =$ 17, 5 and 4~\msun\ for outflows 1, 2 and 3. In the cases of outflows 1 and 2, red and blue lobes are well resolved and the corresponding centers have separations of $\sim$~47\arcsec\ and 81\arcsec, which are larger than APEX HPBW of $\sim$ 20\arcsec. This suggests that the inclinations angles should be small. The length of the blue lobe of outflow 2 from the terminal H$_2$ bow shock to the proposed driving source is $0.9\pm0.3$~pc. However, this parameter depends on the (small) inclination of the outflow with respect to the plane of the sky.

Finally, we propose a simple scenario that puts together the evidence found in this work and foresees a common origin for all shock excited emissions and the two outflows in the region, provided they are all at the same distance. This scheme also suggests the binarity of the exciting source as a natural explanation for the observed distribution of shocked emissions and/or entrained gas, which forms the blue and red-shifted lobes of the two molecular outflows in the region. Thus, this cartoon model highlights the physical connection among optical, NIR and molecular outflows in young stars.

%=========================================================================================
\section*{Acknowledgements}

We especially thank to Dr. Mischa Schirmer for his support and advises with the THELI's reduction package and to Dr. Rodrigo Carrasco, the instrument scientist for GSAOI, for suggestions during the observation process.
We also would like to thank the anonymous referee for his/her thorough revision of the paper and his/her constructive comments and suggestions that improved the manuscript. This work was partially supported by a grant from SeCyT--UNC, Argentina. M.R and H.P.S wish to acknowledge support from CONICYT (Chile) through FONDECYT grants No1140839 and No1190684. M.R acknowledges partial support from the project BASAL PFB-06. H.P.S appreciates the financial support of a postdoctoral fellowship from SeCyT--UNC.

%%%%%%%%%%%%%%%%%%%%%%%%%%%%%%%%%%%%%%%%%%%%%%%%%%
\section*{Data availability}

The data underlying this article are available in Zenodo, at \url{https://doi.org/10.5281/zenodo.3903751}.



\bibliographystyle{mnras}
\bibliography{biblio} 
%%%%%%%%%%%%%%%%%%%%%%%%%%%%%%%%%%%%%%%%%%%%%%%%%%

%%%%%%%%%%%%%%%%% APPENDICES %%%%%%%%%%%%%%%%%%%%%

%\appendix

%\section{Imagenes extras}

% If you want to present additional material which would interrupt the flow of the main paper,
% it can be placed in an Appendix which appears after the list of references.

%%%%%%%%%%%%%%%%%%%%%%%%%%%%%%%%%%%%%%%%%%%%%%%%%%

% Don't change these lines
\bsp	% typesetting comment
\label{lastpage}
\end{document}